\documentclass[amsmath,amsfonts,amssymb,onecolumn,superscriptaddress,showpacs,nofootinbib]{revtex4}

\usepackage{graphicx,psfrag,color}
\usepackage{dcolumn}
\usepackage{bm}
\usepackage{mathbbol,verbatim}
\usepackage{hyperref}

\newcommand{\caligraphic}{\mathcal}

\newcommand{\gge}{\gg}

\begin{document}

\title{From Higgs to pions and Back-  the Unbearable Lightness of a Composite Scalar Boson at 125 GeV in Purely Vectorial Theories}

\author{Shmuel Nussinov}
\affiliation{School of Physics and Astronomy Tel Aviv University, Tel Aviv, Israel}
\affiliation{Institute for Quantum Studies Chapman University, Orange, CA 92866, USA}
\affiliation{Dept. of Physics University of Maryland, College park, Maryland}

\date{\today}

\begin{abstract}
  We argue that the $125 GeV$ ``Higgs'' particle is unlikely to arise as a fermion- antifermion composite if the underlying dynamics is a vectorial gauge theory. The reason is that  the lightest scalar in such theories is heavier than the lightest pseudo-scalar with the mass difference being fixed by the scale of the theory.  LHC searches suggest  that the scale of any new physics, including that of a putative new theory dynamically generating the 125 GeV  ``Higgs'' particle, is relatively high $\sim{(1/2TeV-1TeV)}$. Also the LHC analysis suggests that it is {\it scalar} namely $J^P = 0^+$ rather than pseudo-scalar. Thus it is unlikely that the ``Higgs'' could arise as a composite in such theories- though it will arise in special cases when the underlying binding gauge group is real as a fermion-fermion bound state.

The direct considerations of the various two point functions in the large $N_c$ limit presented below- suggest that massless pseudo-scalars, but not any other anomalously light meson, arise as composites of massless fermions say the massless u and $\bar{d}$ quarks in QCD. These massless pions manifest the spontaneous breaking of the global axial symmetry in QCD with the pions being (pseudo) Nambu Goldstone Bosons. This offers a different insight into SXSB in QCD and most other confining non-abelian gauge vectorial gauge theory. Specifically we consider the euclidean two point functions $F_I|x-y|$ for asymptotic $|x-y|$ expressed as a sum over fermionic paths. We conjecture that for the pseudo-scalar two point function - and for that case only- self retracing paths and closely related paths make in this limit a positive, coherent and dominant contribution, a contribution which evades the generic asymptotic exponential fall-off and allows the lightest pseudoscalars to be massless. The same arguments imply that the scalars are very massive. 
\end{abstract}

\pacs{05.30.-d, 03.67.Pp, 05.30.Pr, 11.15.-q}
\maketitle

\section{Introduction}
 
 No evidence for new physics (NP) beyond the Standard Model(SM) such as supersymmetry, Technicolor, new dimensions, escaping dark matter particles or any other unexpected effect, has been found to-date at the LHC. The scale characterizing new physics such as masses of new particles or that of any new (possibly confining) gauge theory is thus bound from below by $\sim{(1/2TeV-1TeV)}$.

 On the other hand the heroic efforts of the CMS and ATLAS Collaborations yielded a boson $H_0$ at $\sim126 GeV$ most likely with spin/parity $0^+$. While the rates of it's decay into $b$ quark or $\tau$ lepton pairs have not been accurately measured, it may be the last missing element in the SM, the Higgs particle which is needed for a renormalizable theory based on the $SU(2) \otimes U(1)$ gauge group describing the electro-weak (E.W.) interactions. Such a crowning success of the SM further limits NP, such as a massive fourth generation, some right handed neutrinos or other new particles into which the H rapidly decays \cite{Carmi},\cite{Bhupal},\cite{Shrock&Suzuki}, etc. Conversely, deviations from production and/or decay rates expected in the S.M. may suggest NP, etc., hopefully to be confirmed in future LHC runs at higher energy. Finding which alternative is correct is crucial for the future of high energy physics and much effort is presently dedicated to this question.

Here we focus on a somewhat different question: Is the new $H_0$ particle fundamental like the leptons, quarks and the SM gauge bosons, or is it composite? All particles and the above ``fundamental'' point-like fermions and vector bosons of the S.M. may be ``Composite'' in one way or another at very high say Planck scales. Thus specifically we ask if the the compositeness of the new dynamics generating the Higgs as a bound state is  much lower, say TeV- PeV, scales and further if in order achieve this {\it some} fundamental scalar particles should exist at these scales.

 We assume that the $H_0$ is related to the (EW) symmetry breaking so that in dynamical models the $H_0$ is a composite of a fermion and anti-fermion or other objects which carry SU(2)$_L$ $\otimes U(1)$ quantum numbers rather than be a``glue-ball'' analog made of $SU(2)_L \otimes U(1)$ singlet ``gluons''.  To-date all known spin 0 quanta such as compressional phonons, the Cooper pairs in ordinary low temperature super-conductors and the light, pseudo Goldstone QCD pions are {\it not} elementary. Indeed theories with elementary spin 0 bosons like  $\lambda \phi ^4$  or Yukawa models with fundamental (pseudo)-scalars are ``trivial'' \cite{Aizenmann} and have Landau poles related to UV divergences. The self coupling of a $125 GeV$ Higgs in the standard model is relatively small and such difficulties appear only at unreachable energies. Also possible tunneling into another more stable vacuum state takes times far longer than Hubble time. Still, having the first, elementary scalar ever has  profound implications.

 We will argue that in a large class of dynamical EW symmetry breaking schemes with vector like underlying gauge theories with positive Euclidean path integral measures and with intrinsic compositeness scales $~TeV$ or higher where the Higgs boson is supposed to be a fermion-anti-fermion composite, it cannot be as light as the $126 GeV H_0$. Further, if the underlying dynamics are Xiral gauge theories where the above positivity and our arguments fail, the 126 GeV H(0) may not have a well defined (positive) parity or should be accompanied by an almost degenerate pseudo-scalar..  

\section{The General Set-up and Lines of Arguments}

  QCD which underlies strong interactions, confines quarks and spontaneously breaks (SXSB) the $SU(2)_L \otimes SU(2)_R$ global symmetry for mass-less u and d quarks into the diagonal, vector,  iso-spin symmetry via the condensate :$\langle \bar{q}q \rangle = \langle \bar{q}_L q_R \rangle \sim (300 MeV)^3$ where $q=  u,d$. The associated Nambu Goldstone (N.G) bosons are the $\pi^+, \pi^-$ and $\pi^0$.

  E.W. symmetry breaking can be dynamically generated via SXSB in a QCD analog (say Technicolor) gauge theory of a scale about $10^3$ higher. The resulting Goldstone pseudo-scalars form the longitudinal modes of  $W ^+, W^-$ and $Z^0$. Such schemes maintain the global Xiral symmetries which forbid quark and lepton masses. Thus unlike the SM where the Higgs condensate generates all lepton and (bare) quark masses, generating fermionic masses in composite Higgs models requires additional elements such as ``Extended Technicolor''. The latter is  also needed to avoid large FCNC (flavor changing neutral currents) and some incorrect predictions of simple Technicolor which conflict with precision E.W measurements.

Independently of all such details we will argue that the very lightness of a scalar $H_0$ tends to excludes it's compositeness -as a fermion anti-fermion bound state-in any underlying non-abelian confining vectorial theory.  Rather than appeal to the pattern of hadronic masses we wish to understand {\it why} the mass difference between the pseudo-scalar and scalar in QCD is large as the same arguments may apply also to the case at hand.

 In sec III we recall how the Naive Quark Model (N.Q.M) addresses this issue by invoking a very strong attractive``Hyper-fine'' (H.F.) interaction in the $^1S \bar{q}q$ channel (here $u\bar{d}$)  which has the $J^P=0^-$ quantum numbers of the ground state pseudo-scalar ($\pi^+$ for QCD). Similarly for $qq$ (rather than a $\bar{q}q$) the strong H.F. attraction in the $^1S$ {\it scalar} channel generates the relatively light I=0 scalar diquark.

We suggest in sec VI that these features of the H.F. interaction in QCD can be related to the coherent contribution of a subset of fermionic path pairs in the path integral. In particular for the special case of an $SU(2)$ color confining gauge group we can have practically degenerate $\bar{u}d$ pion and the color singlet $ud$ diquark ``baryon''.

In sec IV we review the inequalities between euclidean two point functions which imply mass inequalities. The inequality $m^0(0^+)\le{ m^0(0^-)}$ stating that the lightest I spin non singlet, $0^+$ state is heavier than the lightest pseudo-scalar, is closely related to the Vafa-Witten theorem. The nucleon-pion mass inequality originally due to Weingarten, along with the anomaly matching conditions allows proving SXSB in the Large $N_c$ limit using the arguments first presented by Coleman and Witten.

In sec V we use the inequalities, some basic general features of confining vectorial theories {\it and} the assumption of SXSB, to argue that the splitting between the lightest scalar and the lightest pseudo-scalar is large - of order $\mu$, the scale of the vectorial theory.

In Sec VI we argue that this result may follow, even without assuming SXSB, from direct estimates of the path integral for the pseudo-scalar and scalar two point functions (T.P.F's). These argument suggest that the lightest pseudoscalar- the pions in QCD- are massless when$ m^0(u) =m^0 (d)=0 $, which may eventually provide an alternative proof of SXSB for $N_c\rightarrow\infty$. The qualitative difference we find between the scalar and pseudo-scalar TPF's argues against massless, or light scalars which are fermion anti-fermion composites.

In sec VII we note that LHC limits the scale $\mu$  of a new underlying gauge dynamics invoked  to generate composite 125 GeV scalar Higgs, to be higher than $\sim{(1/2-1)} TeV$ order TeV and that LHC suggests both a spin zero and positive parity $J^P= 0^+$  for the $125 GeV$ particle is most likely scalar ($J^P= 0^+$). We also detail more carefully the extended set of assumptions needed for Sec V and VI which argue against a composite Higgs in most models which do not have elementary scalars to start with.

In Sec VIII we reiterate the caveats that our discussions are limited to theories where the path integral measure is positive. This is not the case for Xiral gauge theories or theories with a large $\theta F \cdot \tilde{F}$ term. Such theories could conceivably generate a composite, light 125 GeV scalar. Positivity does also fails in most SUSY theories \cite{Cvetic} While such theories could conceivably generate a composite, light 125 GeV scalar they nicely accommodate many fundamental scalars and there may be less incentive in this case to have a composite 125 GeV particle in the first place. We also briefly note that the positivity is a relatively robust feature and small $\theta F \cdot \tilde{F}$ and /or $\epsilon\bar{\psi(x)}\gamma_5.A_{\mu}(x)\psi(x)$ may not sufficiently spoil it so as to dramatically change the conclusions drawn.

 Finally sec IX provides a short summary. 

\section{Scalar Pseudo-Scalar Mass Splitting: the NQM Arguments}

The SXSB in QCD transforms the original almost mass-less, bare $u^0 , d^0$ quarks into massive $u$ and $d$ ``constituent'' quarks consisting of the original  point-like quarks and a cloud of gluons and $\bar{q}q$ pairs. The Na\"ive (yet successful!) Quark Model (NQM) treats the low lying hadrons as bound states of constituent quarks of masses $m(u)\sim{m(d)}\sim 350 MeV$, and a strange quark of $m(s)\sim 480 MeV$. A smooth confining potential avoids quark liberation and for slow quarks a non-relativistic treatment is applicable. However in order to``explain'' the very light pseudo-scalars a strong hyper-fine interaction of the chromo-magnetic moments $\mu (i)$ of the quarks has to be invoked.

The moments are taken to be: $\mu(i)=g\lambda(i)/{m(i)}$ with $m(i)$ the constituent quark mass, $\lambda(i)$ its $3\times 3$ color matrix and $g$ is the QCD coupling. With $\mu(i)$ parallel to $s(i)$ with $s(i)$ the quark spins, the interaction between quarks $ q(1), q(2)$ in a baryon or between $q(1), \bar q(2)$ in a meson is proportional to the scalar product (in spin and in color space):
\begin{equation}
 H _{H.F} = - C\mu(1)  \mu(2) (\lambda (1)  \cdot \lambda (2)) (s(1) \cdot s(2)) \delta^3(r)
\label{equation HF}
\end{equation}
 Since $s(1)\cdot s(2)$ is $1/4 , -3/4$ in overall $S=1, S=0$ states, $H_{H.F}$ pushes up the mass of the $\bar{q}(i) q(j)$ triplet (vector) state and pushes down the mass of the singlet (pseudo-scalar) states by three times as much. Fitting the strength C and using first order perturbation yields the correct splitting between the spin $3/2$ decuplet and spin $1/2$ octet of baryons and between the octets of vector and pseudo-scalar mesons \cite{Rujula},\cite{Karliner},\cite{Richard}.

An  important feature for our purpose here is that $H_{H.F.}$ also generates a large mass difference between the lowest scalar and pseudo-scalar: in a $0^+$ state, the $\bar{q}q$ have, due to the negative relative parity of a fermion and anti-fermion, $L=1 $angular momentum. The delta function in $H_{H.F}$ makes it's contribution proportional to the wave function at the origin, which vanishes in the $L=1$ scalar states but not in the $L=0$ pseudo-scalar states where it manifests as a very strong attraction. Note that for a $qq$ system the situation is exactly the opposite- and a strongly bound {\it scalar} $^1S$ state is preferred.

The above ``explanation'' of the $630$ MeV $\rho \pi$ mass difference using the very strong H.F. Attraction does not seem to be viable for many reasons and in particular the non-relativistic treatment should break down when the constituent quarks bind to $m(\pi)\sim 0$. However as the quark and anti-quark overlap and their color fields cancel and larger portions of the constituent quarks clouds disappear. At zero separation the system becomes just two mass-less quarks with the ($\sim$ 700 MeV) sum of constituent masses acting as a huge binding ``justifying'' the strong H.F. interaction \cite{NSh}.

 The fact that the Goldstone pion is  a``collective'' state with many $\bar{q}q$ and yet seems to be a normal $\bar{q}q$ meson with form-factors and cross sections similar to those of other mesons is the well known $\rho-\pi $ puzzle \cite{Pagles} further elaborated on in \cite{NSh}. The path integral arguments for mass-less pions presented  in Sec VI may resolve this puzzle along the lines originally suggested by Pagles\cite{Pagles}. In turn it may justify also some of the applications of the NQM with the strong H.F interactions for baryons.

 \section{Mass Inequalities in Vector-like Theories}

  Inequalities in vector-like theories provide our main tools. We later present a modified approach involving sums over fermionic paths and attempt to directly show that the pseudo-scalar is massless in the limit of vanishing bare quark masses. In this section we briefly review the ``standard''  QCD inequalities and in the next section argue that these inequalities together with the conventional assumptions regarding non-abelian strongly coupled, gauge theories, motivate large masses for all non- pseudo-scalar particles.

  The positive path integral measure in vectorial gauge theories  :
\begin{eqnarray}
 d\mu(A)= D(A_{\mu}(x))\exp[-S(A(x))] \prod_{a=1,2,..N_F} Det(D(A) + m_{a})
\label{equation: measure}
\end{eqnarray}
allows probabilistic Monte-Carlo lattice QCD calculations. $S(A(x))=\int d^4 x [ E^2 +B^2]$ is the Euclidean Y.M. action,
 $D(A)=\gamma_{\mu}.[\partial_{\mu}+iA_{\mu}(x)] $ is the Dirac operator in the background of the $A_{\mu}(x)$ gauge fields and the product extends over $N_F$ fermionic flavors a with masses $m_a$. The determinants are positive since $\gamma_5\gamma_{\mu}\gamma_5=-\gamma_{\mu}$ and: 
\begin{equation}
\gamma_5 D(A) \gamma_5= D^+(A)             
\label{equation: gamma_5conjugation}
\end{equation}   
causing the eigenvalues of the anti-Hermitian Dirac operator to appear in complex pairs $+i\lambda(A)  $ and $ -i \lambda(A)$ so that :
\begin{equation}
  Det[D(A)+m_a] = \prod_\lambda (\lambda(A)^2 + m_a^2)         
\label{equation: Det positivity}
\end{equation}
This pairing and positivity fail in chiral theories where $\gamma_5\psi(\lambda) = +/- \psi(\lambda)$ with $\psi(\lambda)$ the fermionic eigenmode.

\begin{figure}[htbp]
	\centering
		\includegraphics[width=.25\textwidth]{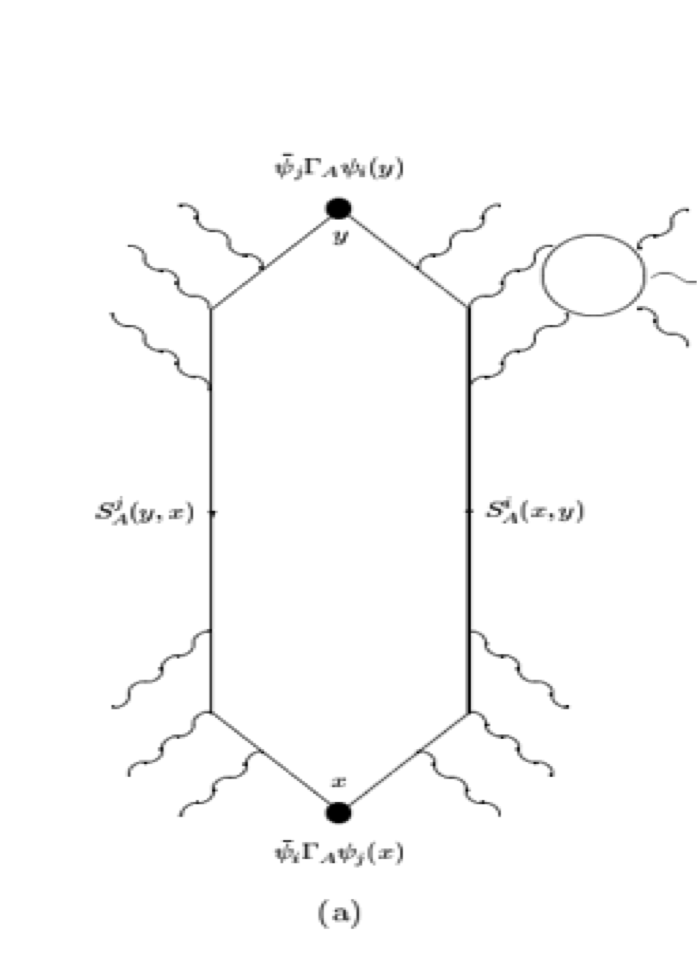}
		\caption{One loop of propagators contributing to the TPF. The wiggly lines indicate the interactions with background gauge fields and the closed fermionic loops correspond to the perturbative expansion of the determinant factors in the measure.}
		\label{slide6}
\end{figure}

Consider the Euclidean two point functions  (TPF's) 
\begin{equation} 
F_I^{ab}(x,y)=  \langle0 |J^{ab}_I(x) J^{ab}_I(y)|\rangle
\label{equation:TPF}
\end{equation}
 of fermionic bilinears are $J^{ab}_I(x)=\bar\psi_a(x) \Gamma_I \psi_b(x)$ with $\Gamma_I; I=1,.., 16$ the elements of the Dirac algebra. It is given by the path integrals :
\begin{equation}
F_I^{ab}(x,y) = \int d\mu(A) tr[\Gamma_I S_a(A_{\mu})(x,y) \Gamma_I S_b (A_{\mu}(y,x)]  
\label{equation: PI for 2Ptf}
\end{equation}
which are pictorially illustrated in Fig.~\ref{slide6}.

The invariance of $F_I(x,y)$ under translations and four dimensional rotations (or Lorentz transformation in the Minkowskian version) which follows from it's definition and the invariance of the vacuum, is ensured by the path integral extending over all translated and rotated version of each background field configuration. This implies that $F_I(x,y)$ depends only on the separation $|(x-y)|$. In the above equation $(a,\bar{b})$ are the flavors of the fermion/anti-fermion, $S_a(A_{\mu})(x,y)$ is the propagator of a fermion of flavor a, from x to y in the presence of a background gauge field $A_{\mu}(z)$. The propagator, a $3x3 \otimes 4x4$ matrix in color and spinor space, is defined by: $[D(A) +m_a)] S_a(x,y) = \delta^4(x-y)$ where we suppressed the color and spinor indices.

 $\gamma_5$ conjugates both $[D(A) +m_a)]$ and it's inverse (the propagator):
\begin {equation}
 \gamma_5 S_{A_{\mu}}(x,y) \gamma _5 = [S_{A_{\mu}}(y,x)]^+,
\label{equation:Propagator conjugation}
\end {equation}
with$ ^+$ indicating conjugation in color and spinor space.

Disconnected annihilation diagrams with an extra fermionic loop and an extra minus sign are absent in the large $N_c$ limit which we will adopt and are absent anyway in bilinears involving fermions (quarks for QCD) of different flavors a and b. If further $m_a=m_b$, or in QCD $m_u=m_d$, and weak/em gauge interactions are switched off, then also $S_a=S_b=S$. Thanks to the $\gamma_5$ conjugation property, the pseudo-scalar two-point function (TPF)  is given by a path integral with positive measure of a  positive integrand: $S^+(x,y)S( x,y)$. This implies that the pseudo- scalar TPF is larger than or equal to all other TPF's \cite{Weingarten},\cite{Vafa1} and in particular:
\begin {equation}
\langle 0| S(0) S(x)|0 \rangle  = F_S(x)   \le  \langle 0|P(0) P(x)|0 \rangle = F_P(x)     
\label {equation: S&P}
\end {equation}

The decomposition of the fermionic propagator: 
\begin{eqnarray}
S_F(x,y;A)= s(x,y;A)I + \gamma_{\mu} v^{\mu}(x,y;A)+ \sigma_{\mu\nu} t^{\mu\nu}(x,y;A)  \nonumber
\\ + \gamma_5 \gamma_{\mu} a^{\mu}(x,y;A)+ \gamma_5 p(x,y;A)
\end{eqnarray}
and the $\gamma_5$ conjugation, imply that the pseudo-scalar TPF is  
\begin{eqnarray}
F_{ps}(x,y)=\int d\mu(A)[|s|^2+|v^{\mu}|^2+|t^{\mu,\nu}|^2 +|a^{\mu}|^2+|p|^2] 
\end{eqnarray}
 were we suppressed the x,y co-ordinates and background field $A_{\mu}$ dependence of all s,..p entries. The four $v^{\mu}$, four $a^{\mu}$ six  $t^{\mu,\nu}$ and s,p correspond to the 16 elements of the $4x4$ spinor matrix in $S(x,y)$. For the scalar TPF eight of the terms reverse sign
\begin{eqnarray}
F_{s}(x,y)=\int d\mu(A)[-|s|^2 +|v^{\mu}|^2-|t^{\mu,\nu}|^2 +|a^{\mu}|^2-|p|^2].
\end{eqnarray} 
These terms are non-zero yielding the strict inequality \cite{N&S} 
\begin{equation}
  F_P (x,y) > F_S (x,y)       
\label{equation: strict PS}  
\end{equation}

All TPF's have spectral representations :  
\begin{eqnarray}
F_I(|x-y|) =\int d\mu^2 \sigma_I(\mu^2)\exp {-(\mu \cdot |x-y|)} .
\label{equation: spectral rep}
\end{eqnarray}
 The currents $J_I$ prescribe the spin-parity of the physical states contributing to the spectral function: $1^-$ for vector current,$0^+$ for scalar current etc. As $|x-y|$ tends to infinity the state(s) of lowest mass $m^0$ dominate and ,up to coefficients and inverse power of $|x-y|$ in pre-factors, the TPF's behave like: $\exp{-[(m ^0_I).|x-y|]}$ .The inequalities between TPF's lead then to reversed inequalities between the masses of the lowest lying states in the corresponding channels. If these are discrete, single particle states, say the $0^- \pi $, the $1^- \rho $ the $1^+  A_1$ and some scalar $a_0$, we find that: $m(\pi)\le{m(\rho)}$ ; $m(\pi)\le{m(A_1)}$ and the mass inequality which is most relevant to the present paper :
\begin{equation}
        m^0(0^+[u,\bar d]) \ge m^0(0^-[u,\bar d])                   
\label {equation: m of S&P}
\end{equation}
stating that the lightest scalar in the $ 0^+[u,\bar d] $ channel is heavier than the lightest pseudo-scalar in the $0^-[u\bar d]$ channel.

 This mass inequality is related to the Vafa-Witten theorem\cite{Vafa1} that the isospin symmetry of QCD (or any global vectorial symmetry in vector like gauge  theories) cannot break spontaneously. To show this let us turn on small, equal, u and d bare masses: $m^0(u)=m^0(d)= m^0$ and turn off electro-magnetism. This ensures that: $ S_u(A_{\mu})(x,y)=S_d(A_{\mu})(x,y) $ allowing the derivation of the above mass inequality. A spontaneous breaking of the exact Isospin symmetry would generate massless scalars; on the other hand the explicit breaking of the axial symmetry by bare quark masses renders the pion a pseudo-Goldstone particle with a small mass violating the above inequality. We therefore conclude that the spontaneous breaking of the global, vectorial Iso-spin) symmetry cannot happen.

En route to the theorem the seminal paper by Vafa and Witten\cite{Vafa1} has the critically important result that apart from overall numerical coefficients the propagator for a fermion of bare mass $m^0$, in any background gauge field is bound by $exp{-[m^0|x-y|]}$:
\begin{equation}
 S^{Delta}_{A_{\mu}}(|x-y|)\le\exp{-[m^0|x-y|]}/{m^0\Delta^4}
 \label{equation: VW bound}
\end{equation}
 Where to avoid anomalously enhanced propagation by spiraling tightly around a large B field directed along x-y and  to ameliorate the gauge non- invariance of S a smearing of the initial and final points x and y over balls of radius $\Delta$ has been performed.

The inequalities imply that the pion is the lightest $ \bar{q} q $ state. In principle flavor singlet glue-balls and specifically the scalar glue-ball  \cite{Muzinich} could be lighter. However the $0^{++}$ glue-ball state is orthogonal to the non-trivial vacuum state with a non-zero $F_{\mu\nu}^2$ condensate and is expected to have a mass of $\sim \mu $ with $\mu$ a characteristic mass gap in QCD or the general non-abelian gauge theory considered.

 Weingarten\cite{Weingarten} derived also the inequality $m_N\ge{m_{\pi}}$. In the large $N_c$ limit \cite{tHooft},\cite{Coleman} the inequality becomes: $m_N > (N_c/2) m_{\pi}$\cite{N&Sat} and can be used to prove SXSB. Matching the axial anomaly at the quark level \cite{'tHooft} requires having a zero energy {\it pole} in the pseudo-scalar channel in the physical spectrum\cite{Frishman}\cite{Coleman2}. This can be achieved by a mass-less pion, by a pair of massless nucleons or by combinations of both. However the tiniest pion mass due to small but non-vanishing bare quark masses
$m^0(u), m^0(d)$ implies infinitely massive nucleons. As in the original argument by Coleman and Witten \cite{Coleman3} this leaves only the massless pion option and proves SXSB in the large $N_c$ limit.
An attempt to prove SXSB for all $N_c$ using $m_N > m_{\pi}$ was made in the review on QCD inequalities \cite{N&L}. Explicit breaking of the axial symmetry via bare quark masses are expected to generate for an other-wise assumed massless nucleon, a small mass {\it linear} in $m^0$. If on the other hand $ m^2 (\pi) \sim  C' \mu (m^0(u) + m^0(d)$ then dialing up the bare quark masses keeping $(m^0(u) + m^0(d)\ll \mu $ will violate $m_N > m_{\pi}$ and SXSB would then follow from considerations of the anomaly as above. The GMOR relation \cite{Gell-Mann} has the above form for the pion mass:
\begin{equation}
      m_{\pi}^2 = \frac{\langle \bar{q} q \rangle}{F_{\pi}^2} ( m^0(u) + m^0(d))     
\label{equation:GMOR}
\end{equation}

  Since``soft pion'' techniques are used in deriving the GMOR relation and such techniques are justified only for Goldstone type pions this argument is circular. Note though that the relation is largely independent of $N_c$ as both $\langle \bar{q} q \rangle$ and $F_{\pi}^2$ are proportional to $N_c$.

 Our arguments in Sec VI do {\it not} provide an independent proof of SXSB in general or in the large $N_c$ limit. Still trying to understand how massless pions arise in a path integral formulation and the qualitative difference between the scalar and pseudoscalar two point functions therein is useful.

\section{A Large Scalar-Pseudo-Scalar Mass difference- an Approach Based on Estimates of Euclidean Two point Functions}

  The above``ordinary'' QCD inequalities order the masses of the lowest lying states with different quantum numbers. However even the strict inequality between the pseudo-scalar and the scalarTPF does {\it not} imply a stricter inequality $m(0^{+}[u,\bar{d}]) >{m(0^{-}[u,\bar{d}])}$ between the masses of the lightest scalar and the lightest pseudo-scalar. The two masses can be equal with the two TPF's having the same exponential fall-off but different pre-factors.

 We want however to show that the mass difference $m(0^{+}[u,\bar{d}]_{0}) - m(0^{-}[u,\bar{d}]_{0})$ is {\it large}, of order $\mu$, the scale of the theory. For this we need additional properties of non-abelian gauge theories, beyond measure positivity, which hold in any vectorial gauge theory including the abelian and weakly interacting Q.E.D. In particular, we need to introduce $\mu$ - the non-perturbative scale of the theory.

The main assumption which we make in both the present and next section and whose proof is one of the millennium problems is that: 'Vectorial, non-abelian pure gauge theories with conformally invariant Lagrangians, generate via non-perturbative effects a mass gap $M_{gap}$. 

This gap serves as an infrared cut-off and provides a finite correlation length scale: $r_0= M_{gap}^{-1}$ between the gauge fields in the vacuum. A stronger version of this assumption which we also adopt is that such Yang-Mills theories confine quarks. At various points we will further assume that confinement is generated via the``inverse Meissner effect'' and is due to very large chromomagnetic fluctuations in the QCD vacuum.

The general  criterion for confinement uses the Wilson loop:
\begin{eqnarray}
 W (C)= \langle tr(P \exp{i g\int_C  A_{\mu}^a \lambda_a}) \rangle \nonumber
 \\= \int [d\mu(A)]tr(P \exp{i g \int_C  A_{\mu}^a \lambda_a})  
\label{equation: Wil}
\end{eqnarray}
where $C$ indicates the closed path along which the gauge field $A_{\mu}^a \lambda_a$ is integrated yielding the ``non-abelian phase'' collected for each gauge field configuration $A_{\mu}$, $P$ indicates path ordering and $\lambda_a ,  a= 1,2,..N_c^2-1$ are the generators of $SU(N_c)$ in the fundamental representation. Finally we average over the gauge fields with the measure $d\mu(A)$ dictated by the Yang-Mills action.

 The Wilson loop $W(C)$ serves as an effective gauge invariant ``order parameter''. Let  $C$ be planar, non self-intersecting and smooth. If it's linear size l is large relative to $r_0$ then $W(C)$ can behave as: $ \sim\exp[-p/r_0] $ with $p$ the perimeter of $C$ or as: $\sim \exp[-A/r_0^2]$ with $A$ the area enclosed by $C$. (For non planar $C$ the area $A$ is defined by the surface of minimal area having $C$ as its boundary) 
It is well known that the area law implies a confining linear potential  $ V(R) \sim \sigma R $ between a static (infinitely heavy) quark Q at $z=-R/2, x=y=0$ and a static anti-quark $\bar Q$ at $z=R/2, x=y=0$. In QCD the coefficient $\sigma = \mu^2$ is $\sim 0.16 GeV^2$ and $\mu \sim 400 MeV$.
\footnote{ 
 W(C) can depend on a host of other parameters. In the original conformal theory, instead of the above l and $l^2$ behaviors, W(C) depends on the shape of C and varies as $\log{l}$ . Also the curve (C) can form a knot with many topological aspects which may reflect in W(C), can be a fractal, etc. For the purpose of our largely qualitative heuristic argument we assume that the finite $r_0$ can be used to smooth C sufficiently and will neglect these issues.
}

  All the above applies to pure Y.M theories. Non-dynamical, heavy fermions of bare masses $M^0_a \gg \mu$ only modify the effective coupling by changing the $\beta$ function and in the path integral formulation contribute the following determinants to the path integral measure:
\begin{equation}
  Det (i D_{A_{\mu}} + M^0_a) = Det( M^0_a) Det (1+ iD_{A_{\mu}}/{M^0_a})     
\label {equation: Heavy Dets}
\end{equation}
which can be perturbatively expanded in $1/{M^0_a}$

However in QCD and in putative vectorial theories which break E.W. symmetry dynamically we need to introduce dynamical fermions of bare masses smaller then the scale $\mu$ or vanishing bare masses. We should emphasize that in vectorial theories- unlike theories which are Xiral and/or have elementary scalars- it is impossible to have massive Dirac fermions bind down to zero mass composites. This traces back to the seminal paper by Vafa and Witten \cite{Vafa1} which showed that in the presence of the gauge fields the fermionic propagators tend to decrease relative to the propagators of the same fermions in the free theory.  Since the latter decay asymptotically as  $\exp{-(m^0_a|x-y|)}$ and $\exp{-(m^0_b|x-y|) }$ we have after averaging the product of the two propagators with the normalized positive path integral measure that all TPFs $F_I(|x-y|)$ decay at least as fast as:$\exp{-[(m^0_a+m^0_b).|x-y|]}$ and therefore the masses of the bound states exceed the sum of the bare masses. This in particular implies that in order to obtain massless Goldstone bosons the fermion and anti-fermions constituting them should both have vanishing bare masses. In the following we will then assume that apart from some heavy fermions the light fermions are essentially massless and contain least one $SU(2)_L$ doublet (like the u,d quarks in QCD) so that we may hope to obtain in principle a massless composite Higgs.

If like the QCD quarks, the dynamical fermions are in the fundamental representations of $SU(N_c)$, then the $E$ fields inside the chromo-electric flux tube between the  $Q$ and $\bar Q $ create $\bar q q $ light quarks. $\bar q , q$ screen the color charge of the heavy $Q, \bar Q$ generating the unconfined $\bar{q} Q$ and $q \bar{Q}$ mesons.As the mass of the quark $q_a$ grows above $\mu$ the amplitude for pair production a la Schwinger and the ensuing screening by a $\bar{q_a}q_a$ is strongly suppressed by $S\sim \exp [-(\pi /2) (m(a)^2/{\mu}^2]$ \cite{CNN}. Even for the charmed quark c of mass $m(c)\sim{1.6 GeV}\sim{ 4\mu}$ we find: $S\sim\exp {-25} \sim 10^{-11}$ so that the screening via $\bar{c}c$ pairs is negligible. 

\vspace{1ex}
 Pair creation of even a single light fermion drastically modifies the dynamics by screening the confining potentials. Also massless Nambu- Goldstone bosons prevent inferring the spectrum of low lying $\bar{u}d $states in the various $J^{PC}$ channels by using the spectral representations of the appropriate TPF's as described in Sec IV above. In the presence of massless pions the asymptotic behavior of the various TPF's is controlled by the $2\pi$ or $3\pi$ thresholds and rather then probing QCD the TPF's will depend only the low energy effective Xiral Lagrangians \cite{Gasser}. Some remnant of the special lightness of the pseudoscalars may still manifest for massless pions via the fall-off of the non pseudo-scalar TPF's as higher inverse powers of $l=|x-y|$ due to the softer two and three massless pion threshold singularities as compared with the massless pion pole.

 This ``technical difficulty'' can be avoided by studying the TPF's at intermediate distances : $ 1/{m_{\pi}} \gg |x-y| \gg 1/\mu$.  In the following we avoid the above complications by adopting the large $N_c$ limit \cite{tHooft} \cite{Coleman} where the TPF's are dominated by just {{\it one}} fermionic loop with no pair creation and screening and the determinants $Det (D_A+m^0_a)$ omitted from the path integral measure.

  When the masses of the dynamical fermions vanish all the dimension-full quantities can be expressed in terms of the single mass scale $\mu$ of the theory. Thus $m^0(gb)$, the mass of the lightest glue-ball or the mass gap in the pure Y.M theory ($\sim 1.5 GeV $ in QCD), the mass of the lightest $1^{--} \rho $ vector meson $\sim$ 0.75 GeV, the $O(GeV)$ mass of the nucleon or (the inverse of) it's size, all have the form : $M_i = c_i\mu $ with $c_i$ dimensionless, in principle calculable, constants. Note that ``the''  QCD scale  $\sim\Lambda($QCD$) \sim 200 MeV $ where the running coupling constant diverges, is defined via the perturbative $\beta $ function and not by the non-perturbative physics of interest so that we use in the following the (square root of) the string tension $\mu\sim 400$ MeV.

The basic assumptions used in the rest of this section are:
\begin{enumerate}
	\item The correlation length between gauge fields in the vacuum are finite and of order $r_0  \sim 1/m^0(glueball)$ with $m^0(glueball)=M_{gap}$
	\item The sizes of hadronic states made of the light quarks  are restricted by the confining linear potential to be  $r_0 \sim 1/{\mu}$.
	\item The large $N_c$ approximation, and
	\item The global axial symmetries are spontaneously broken.
\end{enumerate}

 Assumptions 1\&2 are corollaries of the``Millennium Problem'' and in particular of the assumed confinement. Furthermore for confinement to be relevant in the presence of light dynamical fermions we need to add assumption \# 3 and suppress the extra fermionic/quark loops which reflect production of light dynamical fermion pairs and screening. Since SXSB drives the dynamical Higgs mechanism assumption \#4  is  mandatory if the H(0) 125 GeV scalar is a $\bar{F} F$ composite. Thus we explicitly make this assumption though SXSB was proven for large $N_c$ by Coleman and Witten using the t' Hooft anomaly matching conditions \cite{Coleman3}. Assumptions 1-3 are used also in the next section where we directly motivate massless pions without assuming SXSB.

The SXSB implies that in the limit of $m^0_a =0 $ and vanishing $SU(2)_L \otimes U(1)$ radiative corrections, the resulting Nambu- Goldstone pseudo
-scalars are massless. The QCD mass inequalities imply that the E.M mass difference: $\Delta(\pi)= m^2_{\pi^+}-m^2_{\pi^0}$ is positive \cite{Witten},\cite{Nussinov}. The measured or computed \cite{Das} value: $\Delta \sim 36 (MeV)^2$ is only 0.0025 of the corresponding square of non-perturbative QCD scale $\mu ^2 = (400 $MeV$) ^2$.\footnote{
The above sign of the E.M mass difference is crucial for the correct ``vacuum alignment''. With the $\pi^0$ being exactly massless the opposite sign implies tachyonic $\pi^+$ which would condense in the vacuum and lead to spontaneous violation of charge conservation.  
\label{footnote:Inequalities and V.A}
}

In the dynamical Higgs mechanism one often assumes vanishing ``bare'' masses of the fermions $F_a$ which form the $SU(2)_L \otimes U(1)$ breaking condensate. The all important``radiative effects'' drive the Higgs mechanism. Usually one focuses on how these endow three of the four initially massless $SU(2)_L \otimes U(1)$ gauge bosons with masses $m_W$ and $m_Z$. From the vantage point of  the NG bosons the radiative corrections generate for $h^\pm$,$ h^0$ the masses $ m_W , m_Z$  as they become the longitudinal component of  $W^\pm,Z$ respectively.

General arguments based on ``naturalness''  imply that any particle not protected by a symmetry should acquire a mass of order $\mu$. Unless one invokes a new symmetry, such as the approximate dilatation symmetry in specially-tailored theories with $N_F \sim N_c$ discussed below, the mass of the lowest lying scalar is not protected and should be of order $\mu$. The following ``explanation'' why ground state hadrons made of light quarks in all $J^P$ channels have masses $M_I^0 \sim \mu $ is important for showing a large splitting between the ground state scalar and pseudoscalar and also
helps motivate the mechanism for evading this exponential fall-off for pseudoscalars suggested in the next section.

 As before we address the masses by considering the asymptotic behavior of the various TPF's $F_I(|x-y|)$ when $|x-y|=l=T \to \infty$, and we choose $x=(0,0,0,0)$ and $y=(T,0,0,0)$. The wave function of the lightest meson contributing to the TPF of specific currents $J_I$ can be probed by two additional scalar currents: one acting at time $T/2$ and spatial locations $(0,0,r/2)$ (point z) and the second at the same time but at location $(0,0,-r/2) $ (point w). In the large $N_c$ limit the four point function: $F_I(x,y,z,w) = \langle 0|J_I(x) J_S(z) J_I(y) J_S (w)| 0\rangle $ corresponds to one loop of fermions propagating in the gauge field background:
\begin{equation}
 F_I(x,y,z,w) = \int d\mu(A) tr[\Gamma_I S(A_{\mu}(x,z))  S(A_{\mu}(z,y)) \Gamma_I S^+(A_{\mu})(y,z) S^+(A_{\mu})(z,x)]
\label{equation:FOURPF}
\end{equation}

As $T$ ( and $T/2$) become very large the lowest state in the specific flavor and $J^{P}$ channel considered, dominates. At these and most other intermediate times t, the dependence of  $F_I$ on $|w-z|=r$ is given by the real, positive and rotation symmetric ground state wave-function of that state $\psi^0(r)$. Confinement implies that the spatial extent of the ground state wave-function should be $\sim {r_0=\mu^{-1}}$. Under time translation the ground state picks up the phase $\sim{\exp[i(m^0t)]}$ which for euclidean, imaginary, time becomes an exponential fall-off: $\exp [-m^0 t]$.

 In our original TPF the local current $J_I(x)$ created, at time t=0, a quark pair at the same space point and the time evolution was given in the spectral representation as the superposition of many different exponential fall-offs. We can consider instead a superposition of many $\bar{q} q$  pairs symmetrically created at $t=0$ at many spatial locations $+r/2,-r/2$ with amplitude $\psi^0(r)$. In this case the time evolution will {\it not} change this profile and the system will display a single characteristic exponential fall-off :$ \exp{-(m^0 t)}$.

\begin{figure}[htbp]
	\centering
		\includegraphics[width=.33\textwidth]{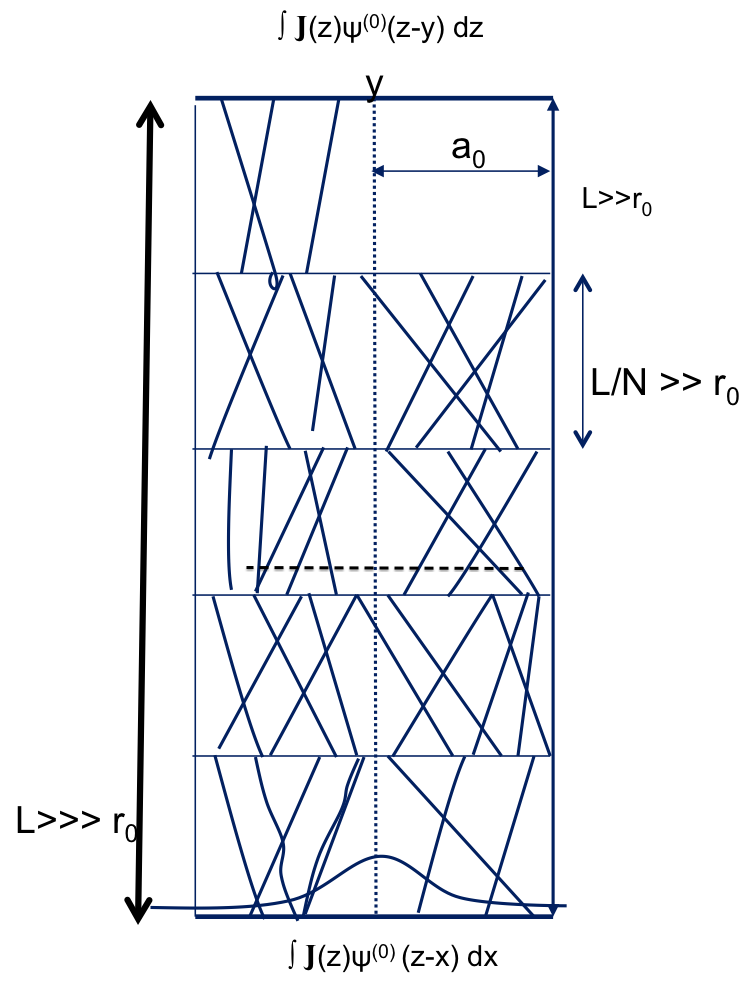}
		\caption{A schematic description of the evolution of a $\bar q q$  system created at $t=0$ (the base of the tall quadrangle) in a  superposition $\sim\psi^0(z-x)$ localized around x at the middle of the base. This initial state is mainly inside the quadrangle of width $a_0$ somewhat bigger than $r_0$ the scale of confinement. The complete time evolution terminates - up to an overall ``phase'' -in  the same wavefunction after a long time $T=L$ at the top of the quadrangle. The evolution is broken up into many time steps of length $L/N\ge {r_0} $ each, with $r_0$ the correlation length of the gauge fields and is indicated by the many straight lines connecting the consecutive time slices. Finally on the third ``rung''  we indicate by the broken line the top of the trapezoid of height $r_0$ and width $\sim a_0 \sim 4r_0$ which represents a coherent evolution element. Note that we drew the vertical $r_0$ as much smaller than the horizontal $a_0$, despite the fact that these lengths are up to a factor which in QCD is: $m^0(gb)/{\mu}\sim 4$ roughly the same}
		\label{slide14}
\end{figure}

In replacing  the creation of a quark pair at the origin: $ J_I(0)= \bar q (0) \Gamma_I q(0) $ by  the superposition:
\begin{equation} 
\int _r \psi^0(r) \bar q (w) \Gamma_I [ P \exp\{i\int_{C_{(w,z)}} g A_{\mu}^a. \lambda_a\}] q(z)
\label{equation:psicreation}
\end{equation}
we maintain gauge invariance by adding the line integral in the square brackets - the ``connection'' between $\bar q(w)$ and $q(z)$. Consider the evolution of the system over a very long time $T \gg r_0 $, which we subdivide into N steps of length $l/N=\delta(t)$. The time evolution over each time step is indicated by the many straight lines in Fig.~\ref{slide14} which represent the propagation of the quark and of the anti-quark at the various possible separations.This evolution and the wave function at all intermediate times is restricted to a ball of radius $a_0=nr_0$ where $n\ge{1}$ a number of order a few is illustrated in the figure by the stripe of width $a_0$.  

 If $\delta(t)=L/N\ge{r_0}$ with $r_0\sim{1/{\mu}}$ the correlation length, then the time evolutions in consecutive segments of size $\delta(t)$ are independent and by time translation invariance contribute the same overall euclidean``phase'' $\exp {(-f)} $ with$f\ge{0}$. The full evolution is then the product: $\exp{[-(m^0l)]} =\exp{[-(Nf)]}= \exp {-[f l /{(r_0)}]} =\exp{-[f \mu.l]}$ and hence $m^0=f\mu$.

Next consider the case when $\delta(t)\le{r_0}$. To find the time evolution over $\delta(t)$ we introduce at time t $P \exp{i\int g A_{\mu}^a \lambda_a}$ where the integral is along the line connecting the quark and the anti-quark, and at time $t+\delta(t)$ the integral in the reversed direction between the new $\bar{q}$ and $q$ locations. We also need the integrals along the``diagonal'' lines of length $\sim \delta(t)$  corresponding to the propagation of $\bar{q}$ and $q$ between the two times. The four lines then encircle a tiny trapezoid of height $\delta(t)$ and upper and lower bases also of size $\sim a_{0}\sim r_0$. The Wilson loop around that little trapezoid generates the time evolution we need. To embed it into the wave function of interest we need to multiply each such trapezoidal Wilson loop by $\psi^0[r(t)]\psi^0[r(t+\delta(t))]$ with r(t) the $\bar{q} q$ separation at time t, and integrate over $r(t)$. The normalization  $\int{[\psi^0(r(t))]^2}=1$ and $\int{[\psi^0(r(t+\delta(t)))]^2}=1$ and the Schwartz inequality then imply that this embedding further decreases the factor associated with the time evolution, making the following argument even stronger.

 To find the time evolution due to this elementary trapezoid we recall that the likely mechanism for confinement\cite{Mandelstam},\cite{'t Hooft 2},\cite{Aharonov} postulates large chromomagnetic fluctuations in the QCD vacuum $ \langle B^2\rangle \sim \mu^4$. Since $\delta(t)$ is smaller then $r_0$ both the height and width of our trapezoid are smaller or equal to the correlation length and the B field of size $B=(B^2)^{1/2}=\mu^2$ is therefore {\it coherent} over this small trapezoid of area $A \sim {r_0.\delta(t)} $. The B field flux through the trapezoid is then $\Phi\sim{Br_0\delta(t)}=\delta(t)/{r_0}$, where we used $r_0=1/{\mu}$. Using the analogy with abelian QED the Wilson loop then supplies a ``phase'' of $(\delta(t)/{r_0})g$ with $g$ the coupling at these scales which effectively translates into an $\exp{-[\delta(t)/{r_0}]g}$ suppression factor.
\footnote{
 Strictly what we obtain is indeed an oscillation phase and to translate it into an exponential fall-off we need to average over it with appropriate weighting as done e.g. in \cite{Aharonov} for deriving the area law.
}

 We considered above the two opposite cases of $\delta(t)\ge{r_0}$ and $\delta(t)\le{r_0}$. We did this so that the evolution along neighboring $\delta(t)$ time slices will be independent in the first case and the B fields will be coherent over the trapezoid of height $\delta(t)$ (and base $\sim{r_0}$) in the second case. To take advantage of both features we will take the common limit of $\delta(t)=r_0$. Assuming that this will only slightly modify the above two arguments regarding the time evolution along the longer/shorter time slices the evolution becomes $\exp{-[g.\delta(t)/{r_0}=\exp-[g]}$. The running coupling $\alpha_s$ diverges at $\Lambda(QCD)$ so that at the scale $\mu\sim 2\Lambda(QCD)$ , $g_s=(4\pi \alpha_s)^{1/2}\ge{1}$. This then yields a suppression by at least $e^{-1}$ over $\delta(t)=r_0$ and by $e^{-\mu.T}$ over the long time T. This then implies that $m^0\ge{\mu}$. 
\footnote{
  In the above we considered the distance between the quark and anti-quark which confinement limits to be of order $r_0$ and neglected  the translational motion of the system as a whole. While the latter involves no dynamics, the energy of the system becomes it's mass only when the momentum P associated with this motion vanishes. Formally this is achieved by superposing :$\int d^3 X \exp{(iPX)}[\Psi^0]$ the wave functions of the system localized at the origin and its shifts via the translation operator $\exp(iPX)$ by all possible values of X. This happens automatically for the local TPF's $F_I[x,y]=F_I[(0,0,0,0),(l,0,0,0)] $: The intermediate states obtained after some long (yet shorter than l) time of evolution are comprised predominantly by the state where in order to minimize the total energy also have overall translational momenta $P$ was forced to vanish.
}

 Adding Fermions tends to increase $r_0$ or decrease $\mu$ but the modification are small in the large $N_c$ limit. Even if we increase the number of light dynamical fermions to $ N_F \sim N_c$  so as to approach an infrared zero of the $\beta$ function, the above argument may still be relevant. All that we used beyond the basic assumption of magnetic field fluctuations of order$ B^2 \sim {\mu^4}$ with $\mu\sim (r_0)^{-1}$ and that at this non-perturbative scale the strong coupling constant is $\ge {1}$.
\bigskip

\section{Estimating two Point Functions by Summing over Fermionic Paths}

 The above arguments suggest that in confining vectorial gauge theories even the lowest lying Fermion antifermion composite states in the various $J^{P}$ channels have masses of order of the scale of the theory $\mu$. Clearly the very light pseudo-scalar pseudo-Nambu-Goldstone bosons associated with the spontaneous breaking of the global axial symmetry are an exception to this rule. How does this happen? What is the mechanism which allows the pseudo-scalar TPF  to avoid the generic exponential $\exp{-(\mu T)}$ fall-off . In this section we suggest an answer this question.

 The key is that the exponential suppression of the various TPF's is due to the strong interactions. Indeed this suppression was ``derived'' above by using the  fluctuating phases due to the strong chromo-magnetic fields in the vacuum of the theory induced by the strong non-perturbative interactions. In the free $g_s=0$ theory all TPFs $F_I(|x-y|)$ in the $u\bar{d}$ channel with massless u and d quarks behave as :$~C_I(|x-y|)^{a_I} $ when
 $|x-y|\to {\infty}$ with I dependent prefactors and inverse powers of $|x-y|$ and have no exponential fall-off which is only due to the strongly interacting phases.

 This suggests that the mechanism generating a pseudo-scalar two point function $F_P(|x-y|)$ with {\it no} exponential asymptotic fall-off and massless pions should involve a particular coherent time evolution with no fluctuating phases and/or flipping of signs of real factors. To see how such a coherent evolution can arise we need a different variant of the path integrals where we sum over fermionic paths. We find that self retracting pairs of paths make  positive coherent contributions to the PS TPF but {\it not} to any other TPF. We conjecture that the contribution these paths (augmented by renormalizations like insertions of side chains which preserve the positivity) evades the exponential fall-off of $F_P(|x-y|)$.

 Thus we suggest that massless pions, or more generally massless pseudo-scalar composites of mass-less fermion anti-fermion $F_a\bar{F}_b$ in underlying confining gauge theories, follow directly from the behavior of the corresponding two point pseudo-scalar function $F_P(|x-y|)$ at large distance. Clearly this is consistent with and follows from SXSB manifesting via the $\langle \bar{q}q\rangle$ condensate in the vacuum since the Goldstone modes correspond to small Xiral rotations of this condensate. Conversely unless an effective Lagrangian with no non-derivative couplings describes the low energy physics of the massless bosons, exhausting all aspects of SXSB, the theory has severe unexpected infra-red (IR) divergences.

The following three differences seem to favor the approach which we suggest here to proving SXSB by showing that $F_P(|x-y|)$ has no exponential fall-off and hence the pions are massless, over the conventional approach of proving the existence of a $\langle \bar{q}q\rangle$ condensate.

\begin{description}
\item[A.] The pion's mass addressed here is the ground state energy in a particular channel with $\bar{u}d$ (or $\bar{F_a}F_b$) flavor and $ 0^- $ spin-parity. This allows using variational arguments which cannot be as readily applied when discussing the order parameter $\langle \bar{q}q\rangle$.

\item[B.] The self retracing fermionic and related paths are chosen here by the positivity of their contribution to the pseudo-scalar two point function which may  allow putting rigorous lower bound on the value of the PS TPF. 

\item[C.] The two points x and y singled out here along the closed fermionic loop allow using the distance $|x-y|=l$ as an important additional control parameter.
\end{description}

  In the strong coupling limit on the lattice only self-retracing fermionic configurations survive by avoiding excessive oscillating non abelian Wilson loop phases. Thus the strong coupling limit featured in earlier efforts to prove SXSB used one loop diagrams shrunk into self retracing parts. The interplay between the present and these works will be an important theme in our discussions

Due to many reasons which will become clear in the following we do {\it not} prove SXSB in QCD and similar Vectorial gauge theories with massless fermions. While several heuristic arguments tend to support our conjecture that the self-retracing fermionic and related paths account for massless pions, much more effort and far more careful analytic treatment is required before a proof obtains or one will find that our conjecture is wrong.

The sum over paths is equivalent to the ``Hopping parameter'' expansion \cite{Creutz} and is least convergent in the limit considered here where the bare quark masses approach zero. Further to maintain exact Xiral symmetry, crucial for the massless pions, and still avoid fermion doubling we need  ``overlap'' or ``domain wall''  fermions \cite{Kaplan1} ,\cite{Narayanan}\cite{Neuberger},\cite{Furman}. While we ignored these important issues, the positivity of the contribution to the pseudo-scalar two point function of the SR paths- the microscopic path by path version of the positivity of the integrand in the standard path integral underlying the``standard'' QCD inequalities- is likely to survive. Being the key to most of the following we hope that the qualitatively features it suggests are indeed correct.

 Our main motivation for considering this approach to SXSB in QCD is not to prove SXSB ( which in the large $N_c$ limit considered has already been proved by Coleman and Witten\cite{Coleman3} but the new 125 GeV scalar discovered at the LHC. If in any vectorial theory the lowest pseudo-scalar bound states of massless fermion and anti-fermion are massless whereas the scalars have masses of the order of the scale of the theory then such theories are unlikely likely to generate a PGNB 125 scalar. For this we do not have to show that the $\bar{F}_aF_b$ pseudo-scalars are massless. Rather it suffices to show that in the above theories mass difference between the lightest scalar and the lightest pseudo-scalar which are made of massless fermion-antifermion is substantial- of order $\mu$ which for the case at hand is $O(TeV)$

 In most formal treatments and also in numerical Monte-Carlo simulations the full fermionic propagators in a given background field are used and the gauge field configurations are summed over only at the end. This procedure is natural as the Lagrangian is bilinear in the fermions. Here we use the reverse order of first summing over all gauge configurations for a given fermionic path P of a u quark going from x to y and a path P' of $\bar{d}$ from y to x, and then performing the sum over all  pairs of such paths.

\begin{figure}[htbp]
	\centering
	\includegraphics[width=\textwidth]{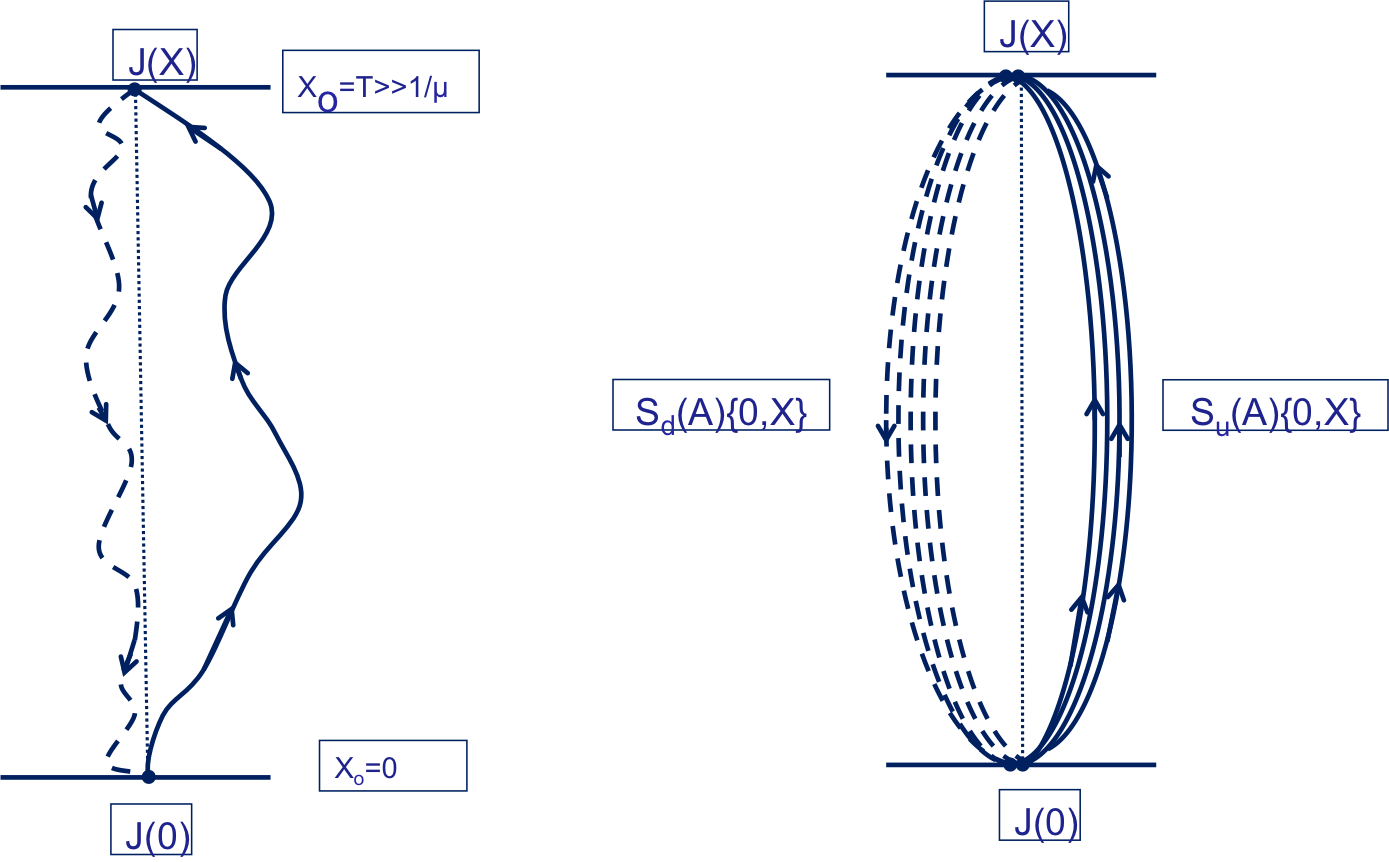}
	\caption{The decomposition of the fermionic propagators into a double sum closed loops made of pairs of up-going $u$ quark paths $P$  and down-ward going $d$ anti-quark paths $P'$ denoted by full and broken lines respectively.}
	\label{slide15}
\end{figure}

As illustrated in Fig~\ref{slide15}
we break the complete u or $\bar{d}$ propagators between points x and y into the contribution of fermionic paths and pair them off into a double sum over closed loops made of the path P of an upward moving u quark and the reversed path P' of a down-ward moving $\bar{d}$ quark. After summing over the gauge configurations with the appropriate weight, which includes the determinants for quarks other than the u and d quarks of interest, we obtain for each pair of fermion paths a gauge invariant Wilson Loop``color'' factor $C_{color}( P,P')$.

We note that the sum over the paths of the massless $u$ and $d$ quarks (which also serve as Alias for the $\bar{F}_a$ and $F_b$ in the general case of interest) is not modified via Pauli blocking or possible annihilation due to loops of the other, non u or d, light flavors and these loops manifest only via the above determinants. This is not the case for loops of the u and d quarks of interest which do affect the sum over paths considered here via Pauli blocking.  All loops are avoided here simply by using the large N(c) limit.

To better control the path summation and understand the spinor factor $C_{spinor}(P,P')$ associated with each loop we discretize the paths using a four dimensional simple cubic lattice with vertices labeled by $n$ with links $ +/- \mu, \mu=1..4 $ emanating from each vertex along the four axes.  
Paths $P$ connecting $x$ and $y$ consist of contiguous links starting at $x$ and terminating at $y$. The  contribution $s(P)$ of each path to the full fermion propagator $ S_F(x,y;A) $ is the ordered product over the links of the gauge group connections $U(n,n+\mu)$ and of the spinorial parts
 $ a+b\gamma_{\mu}$ associated with each link:

 \begin{eqnarray}
 s(P) = \prod_{n,\mu} U( n, n+\mu) (a + b \gamma_{\mu})\label{equation:s(P) contribution}
 \end{eqnarray}

 The real, positive coefficients a and b in the spinor factor are fixed by the bare masses of the quarks, the lattice spacing and by the scheme used for putting the fermions on the lattice. For Wilson fermions $a=b$ and for Kogut-Susskind ``Naive'' fermions we have only the $b\gamma_{\mu}$ part which corresponds to the $(\Delta x)_{\mu} \gamma^{\mu} $ in the numerator of the configuration space massless free fermion propagator. Thus the $\gamma_{\mu}$ is aligned with the link traversed and, in particular, reverses sign when $\mu$ is reversed. The reversed path $\ bar P' $ of the anti-fermion going from from y to x then contributes: 
\begin{eqnarray}
s(\bar P') = \prod_{n',\mu'} U( n', n-\mu') (a - b \gamma_{\mu' }) 
\end{eqnarray}

 Together the paths $( P; \bar P')$  form a loop anchored at the two points x and y. The $P\bar{P'}$ loop makes the following contribution  to the TPF's $F_I ( x,y)$:
 
\begin{eqnarray}
s(\bar P';P) = Tr_{spin,color} [ \Gamma_I s(\bar P') \Gamma_I s(P)] = C_{spin} . C_{color}\nonumber
\\ =   Tr _{spin} [\Gamma_I \Pi_{n',\mu'} (a-b\gamma_{\mu'}) \Gamma_I \Pi_{n,\mu} (a + b \gamma_{\mu })] \nonumber
\\ \times  [Tr_{color}  [\Pi _{n',\mu'} U( n', n-\mu')  \Pi_{n,\mu} U( n, n+\mu)].
    \end{eqnarray}
with  $\Gamma_I$ the elements of the Dirac algebra.

 The $\bar{P}' P$ contribution separates into a product of the spinor term and color term given by the first spinor trace and the second color trace above, respectively. The second trace is the gauge invariant Wilson loop ``phase'' associated with the closed $\bar{P}' P$ loop. It reflects the gauge dynamics and is independent of which correlator (scalar, pseudo-scalar etc) is being considered. The distinction between those is encoded in the first trace.

 The subset of self retracing paths (SRP's) consists of loops where the path of the anti-fermion from $y$ to $x$ retraces in the opposite direction the path of the fermion from $x$ to $y$. Since $ [U( n, n+\mu )] ^+ = U ( n, n-\mu)$, the color factor $C_{color}=\Pi_{n,\mu}[U( n, n+\mu)^+U(n,n+\mu)]=1$, for the contribution of the SRPs to all TPFs.

The key observation is that for the pseudo-scalar TPF (and for this case only!) where $\Gamma_I = \gamma_5$ also the spinor factor $C_{spin}$ is positive for any self-retracing pair of paths. This follows from: $\gamma_5 \gamma_{\mu} \gamma_5 = -\gamma_{\mu}$ and allows the ``microscopic'' path by path version of the general $\gamma_5$ conjugation property underlying the ordinary QCD inequalities as in Eq.~\ref{equation: gamma_5conjugation}. Commuting the $\gamma_5$  through the product  of $\gamma$s corresponding to the links  $(n',n'-\mu')$ in the path $\bar P' $ inside  s(P') above , this product  becomes:$ \gamma_5 s(P') \gamma_5 = \prod_{n',\mu'} (a + b \gamma_{\mu})$.
 For the self retracing paths (SRP)'s $ P'= P$, $n'=n$ and $\mu'=\mu$ and each SR path then makes a positive contribution to the pseudo-scalar two point function ( PS TPF ): $\Pi_{n,\mu} (a+b)^2)$  where the product extends over all the links in the (SR) path considered.

 Thus we see that having the two paths P and P' coalesce makes both the gauge dependent and spin dependent terms simultaneously positive. We suggest that this feature underlies the unusual strong H.F. attraction in the naive quark model in the pseudo-scalar $^1S$ channel that was postulated in order to obtain in that Naive framework massless pions.

 The state of affairs is very different for the $^3P\ 0^+$ scalar $\bar{F} F$ channel in the NQM and more importantly for our present purpose, for the 
$F_S(|x-y|)$ the scalar TPF. Replacing the $\gamma_5$ factors inserted at $x$ and $y$ by $4X4$ unit matrices appropriate for the scalar TPF, omits the above sign flip. Each SR path then contributes : $\Pi_{n,\mu} (a^2 -b^2)$  to the scalar TPF. This should be compared with the positive and in general larger corresponding product $\Pi_{n,\mu} (a+b)^2)$ for the SRP contribution to $F_P(|x-y|)$ the PS two point function.

For general positive a and b the ratio $r=(a^2-b^2)/{(a+b)^2}$ is smaller than unity and hence the contribution of any single SRP to the scalar TPF $F_S(|x-y|)$ is smaller than the contribution of the same SRP to $F_P(|x-y|)$ by $r^N$ where N is the total \# of links on the SRP. Clearly $N\ge{L}$ where $L=l/{\delta}$ is the distance $l$, in lattice spacing $\delta$ units, between x and y. Thus we have the following upper-bound on the ratio of the Scalar and pseudoscalar TPF's 
\begin{equation}
   F^*_S(|x-y|)/{[F^*_P(|x-y|)]} \le{r^L} = \exp{-(\Delta M).(|x-y|)} 
\end{equation}
 where $F^*_I $ will denotes the contribution of the self retracing paths to the TPF $F_I(|x-y|)$ of interest and $\Delta(M)=|\ln(r)|/{\delta}$ is the large mass difference between the scalar and pseudoscalar TPF's which would arise if both were dominated by the contribution of the SRP's.

The most dramatic difference occurs for Wilson fermions:$a=b$ implies $r=0$ and $F^*_S$ strictly vanishes. The case of a=0 on which we will mainly focus in the following is more subtle. Now $r=-1$ and the contributions of the SRPs of length N  to $F_S$ and $F_P$ simply differ by a $(-1)^N$ factor.

 However we next show what, following Oscar Wilde, we call the ``Importance of being positive'', namely that the oscillatory nature of the contribution of the SRP's to $F_S|x-y|$ effectively nullifies them also in the a=0 case. Thus if these contribution evade the exponential decay of the pseudo-scalar TPF (or generate the strong H.F attraction which in the NQM binds the lightest pseudo-scalars to zero) then both features will be absent in the case of $F_S|x-y|$ and in the $^3P$ channel and the corresponding lightest scalar will be heavy.

The following observation which we make in passing is relevant for QCD and even more so for putative different models for Higgs compositeness. If instead of a $\bar{q}q$ system (or a general fermion-antifermion $\bar{F}F$), we consider a $qq$ or $FF$ bound system in underlying vectorial gauge models then the arguments above can be reversed. The self retracing paths can make positive coherent contributions to the {\it scalar} TPF rather than to the pseudo-scalar PF as above. The reason is that the links in the fermionic paths connecting $x$ and $y$ are now parallel and $\gamma_{\mu}^2=1$. The condition for this is that the product of the $U$ matrices be real and coherent. This is readily achieved for real representations $U=U^+$ as is e.g. the case for $SU(2)_c$ (or $O(3)$) where unlike the case of SU(3) the spinorial representation is in fact self conjugate $1/2=\bar{1/2}$. In this case the $0^+$ $ud$ di-quark composite ( which is a boson like the Cooper pairs in ordinary superconductivity) is degenerate with the ordinary $0^- u\bar{d}$ pion. In an analogous theory at a higher scale with appropriate choice of (pseudo-real) confining groups and flavor groups we could use the bosonic di-fermion as a composite lights scalar Higgs.

 Returning to the main discussion, for $a=0$ corresponding to massless naive fermions the self retracing paths contribute $\Pi_{n,\mu}{(b^2)}$ and  $(-1)^ N \Pi_{n,\mu}(b^2)$,where N is the total number of links on the path, to the pseudoscalar (scalar) TPF respectively.
  As above we choose $x=(0,0,0,0)$ and $y=(l,0,0,0)$ and define the integer $L=l/{\delta}$ with $\delta$ the lattice spacing.

 The need to balance ``moves'' along a $+x$ link direction with moves along the $-x$ direction etc. implies that the number of links $N$ on any path connecting $x=(0,0,0,0) $ and $y=(L\delta,0,0,0)$ is even/odd if $L$ is even/odd. Thus the contribution of the the SRP's to $F_S(|x-y)|)$ which is 
$\Sigma_N{(-b^2)^N}$ changes sign every-time that l increases by one lattice unit a truly pathological situation which needs to be rectified.
 Consider next a box centered at $(0,0,0,0)$ ,of size $2d =2n\delta$ with $n\ll{L}$ so that $d\ll{|x-y|}$. This box contains $(2n)^4$ lattice points $ (m_1\delta, m_2\delta, m_3\delta, m_4\delta)$ where $-n\le {m_i}\le {n}$. If x varies over the above ${2n}^4$ points inside the box, the parity of $N$, the number of links connecting $x$ and $y$, is the same as that of  $L+\Sigma {m_i}$ and $F_S(|x-y|)$ changes sign each time we move from one point inside the box to any of it's eight nearest neighbors.

While the lattice introduce an anisotropy the physical TPF's $F_I((|x-y|)$ should depend only on $|x-y|$ and not on the choice of points x and y. 
\footnote{
 An apparent loss of rotation symmetry of the potential between heavy quarks $Q$ and $\bar{Q}$ in strong coupling is a striking lattice artifact \cite{Pearson}. The calculation with the strong coupling lattice Hamiltonian \cite{Creutz} implies that when the quarks are on the $x$ (or $y$ or $z$) axis then the resulting linear confining potential is $V=\sigma R$ with $R$ the distance between $Q$ and $\bar{Q}$ .However for $Q$ at the origin and the $\bar{Q}$ anywhere R is replaced by the minimal distance in the ``Taxi-cab'' metric: $(|X|+|Y|+Z|)$ which for $X=Y, Z=0$ is $2^{1/2}R$ rather than R. The resolution of this utilized the degeneracy of these paths which increases like $c^L$ with $L=R/{\delta}$. In terms of an effective free energy we retrieve the correct potential which is $V(1,2)=\sigma.R(1,2)$ where $R(1,2)$ is the actual Euclidean distance between the quark and anti-quark .
}

Instead of the original $F_I(x-y)$ we can (and for the case of the scalar should!) then use ``regularized'' TPF's defined as the averages when one of the points x or y or both are smeared over a small neighboring region of diameter 2d. For $d\ll{|x-y|}$ the regularized TPF should be same as the original. We can readily verify that this indeed is the case for $F_P^{*} (,x;y)$, the part of the pseudo-scalar two point function which is the sum of the positive contributions of the self retracing paths so that: $F_{reg,P}^{*} (x;y) \sim {F_{P}^{*} (x;y)}$. Thus it leaves open the possibility that for very large distances the PS TPF will indeed be dominated by the contributions of the SRP's which is a part of our conjecture.

 For the case of the scalar TPF however we find that in the sum defining the regularized version    
\begin{equation}
F_{S,reg}(x, y)=\Sigma_{m_1,m_2,m_3,m_4} F_S[(m_1\delta, m_2\delta, m_3\delta, m_4\delta); (L\delta,0,0,0)]/{(2n)^4}
 \label{equation S,reg}
\end{equation}
there is almost a complete cancelation of the contributions of the SRP's to the regularized scalar two point functions originating at even and at odd points even if we use $r_0=\mu^{-1}$ the correlation length in the theory as a natural lattice spacing. Indeed to probe masses $m^0_I\ll{\mu}$ we need the $F_I$ at distances $|x-y|=l\gge {r_0}$ in which case the relative change of l and of $F_I$ when moving between neighboring points is very small and the cancellation due the sign flip of $F_I$ between these points is indeed large.This then rules out the possibility that the contribution of the self retracing paths to $F_S(x-y)$ can- as we argue that it does for the pseudoscalars- generate massless scalars.

 The self-retracing paths contribute the diagonal $P=P'$ part in the double sums: $F_I(|x-y|)=\Sigma_P \Sigma_{P'}[s(P) s(P')]$ for all TPF's. Even when augmented by self-retracing side loops as described below the self-retracing paths are a tiny fraction of all PP' loops. Can the positivity and coherence of the contribution of these paths allow them to dominate the Pseudo-scalar two point function ? and if so does it evade the generic exponential fall off?.

The first question is encountered and often affirmatively answered in many cases. Thus consider a very large set $A= a(1), a(2),..., a(\caligraphic{N})$ and sum over all $(\caligraphic{N})^2$ pairwise products $I=\Sigma{a(i)^*.a(j)}$. If $a(i)$'s are real/ complex numbers with random signs/phases then the double sum is well represented by the diagonal sum: $I=I_d=\Sigma_i{a(i)^*a(i)}$. This holds also for the more general case when the a(i)'s are vectors depending on a randomizing set of parameters which are integrated over. In the case at hand $a_ia_j^*$ is the contribution to $F_{PS}(|x-y|)$ of a  a wider set of closed PP' loops and the diagonal $a_ia_i^*$ corresponds to the self retracing $u$and $\bar{d}$ contributing positively to $F_{PS}$.

Preponderance of self retracing paths occurs also in the condensed matter context of ``weak localization''.  Electrons entering a random medium
can reflect in a direction exactly opposite to that of the incident beam after scattering from impurities. In this case the subset of self-retracing paths where, by time reversal invariance, the phases picked up in the first and second traversals exactly cancel - dominate (see e.g. \cite{Varma}). Also \cite{Altshuler} argued that the density of states near the Fermi-Surface is then modified a feature which may be an analog of the Banks-Casher criterion for SXSB.

\begin{figure}[tbhp]
	\centering
		\includegraphics[height=.3\textwidth]{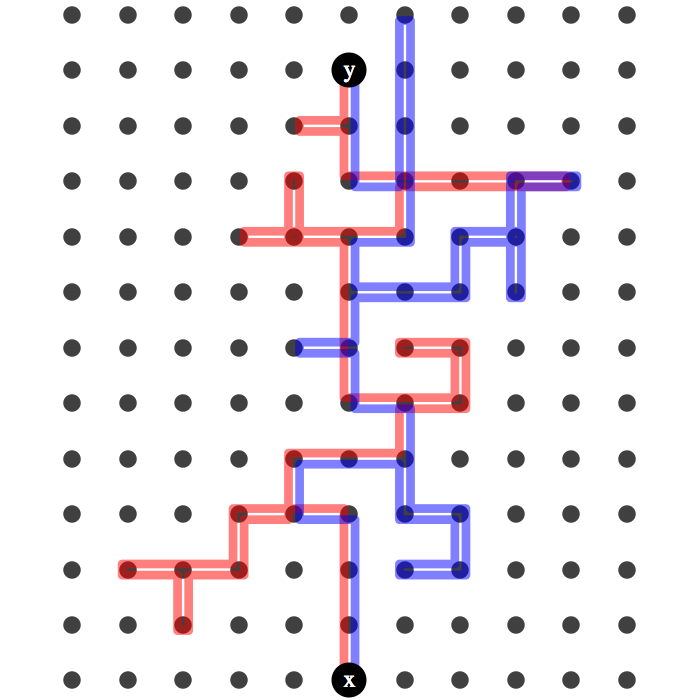}
		\caption{The augmented set of paths in the strong coupling limit. It consists of the original SRP -the trunk made of congruent $u$ links and  $\bar{d}$ links with side branches made of self-retracing paths of u quarks and separately of the d quarks. The trunk connects albeit after some meandering the two special points x and y where we have the $\gamma_5$ insertions whereas the u or d quark lines forming the side chains just leave the trunk and in a self retracing manner return to the same point.
				\label{trunk-and-side-branches}}
\end{figure}

  The random ``phases'' of the Wilson loop , i.e. the  $C_{color}$ factor above generated by the large fluctuations of the color B fields which thread the loop PP', are evaded by adopting the strong coupling limit. This yields a more general set of loops with zero area where sections of the main self retracing path connecting x and y with counter-moving u and bar d quarks, which we will refer to as``the Trunk'', are separated by branches and even complete trees made of self retracing u quarks {\it or} of d quarks. [See Fig.~\ref{trunk-and-side-branches}]. 

Configuration space diagrams reminiscent of the self retracing fermionic paths appeared in several attempts to prove SXSB. Arguing about helicity flips at classical turning points of quark trajectories, A. Casher suggested \cite{Casher} that SXSB follows from binding of mass-less quarks. Later Banks and Casher (B\&C) argued that in the strong coupling limit the vacuum diagrams of one closed fermionic loop generate a non-vanishing $\langle \bar{q}q \rangle $ condensate and SXSB ensues. Similar attempts were due to \cite{Kluberg},\cite{Amati},\cite{Kawamoto}, and \cite{Vafa1}. We refer mainly to B\&C \cite{Banks} who also formulated the criterion for SXSB, that the spectral density of the eigenvalues of the Dirac operator near $\lambda=0$ is the same as that of the free Dirac operator in one dimension.

 In the strong coupling limit \cite{Creutz} adopted in \cite{Banks}, the one loop diagrams ``shrink'' to tree like structures with branches consisting of self-retracing sections similar to the self retracing $u$ and $\bar{d}$ paths considered here. Our choice of the latter among all the PP' closed loops was dictated by the positive and coherent contribution of the self-retracing paths to the pseudo-scalar TPF rather than by the strong coupling limit. Still we find that our original set of SR paths is a subset of the broader class chosen by the strong coupling limit. The latter includes not only of paths of a u quark propagating from x to y followed by the reversed self retracing path of the $\bar{d}$ quark from y to x. Rather, as indicated in Fig.~\ref{trunk-and-side-branches} we can distort the the path of the u quark and independently the path of the d quark at any point of the original $u \bar{d}$ ``Trunk'' and have shrunk side chains or ``trees'' made of pure u or d self retracing quark lines grow at any point on the original joint $u\bar{d}$ self retracing trunk.

 Cutting the long, tortuous, vacuum diagrams considered in \cite{Banks}, reveals at any given ``time''  many $\bar q-q$ pairs in the vacuum with a density
 $\sim\langle \bar{q}q \rangle$. An argument that B\&C attribute to the present author suggests the GMOR relation: $m(\pi)^2\approx m^0\mu$ with $m^0$ the average bare quark mass. Indeed the double commutator: $[Q_5,[Q_5, \bar q q]]$ used in deriving the GMOR relation,\cite{Gell-Mann} inserts two zero momentum pionic ($\gamma_5 \tau$) vertices along the closed fermionic loop converting it for each inserted pair into a pseudo-scalar propagator with zero four momentum of the form described above. Specifically in order to have the above non-vanishing VeV and spontaneous mass generation B\&C require that a finite fraction of the one loop diagrams consist of arbitrarily long trees.  A $\gamma_5.\tau^+$ and a $\gamma_5.\tau^-$ insertions can then be made in a closed loop of a $u$ or a $d$ quark at arbitrarily distant points x and y generating the desired exponentially unsuppressed pseudo-scalar two point function which in turn implies massless pions.

In the preceding paragraph we motivated augmenting our original self retracing $u\bar{d}$ paths by including the shrunk side chains so as to obtain the full  strong coupling set of paths and thereby connect our effort to obtain massless pions via the non-exponential fall-off of the PS TPF with those directed to show that $\langle \bar{q}q \rangle$. To the extent that these earlier efforts have indeed proven SXSB then also ours will.

 We next point  another rational for adding the side chains anchored at points along the $\bar{d}$ or the $u$ quark lines. The insertions of these loops or more generally of any non-perturbative self energy part capable of generating the desired VeV, effectively endows the very light u and d  quarks with masses of order $\mu$ the scale of the theory. The resulting 'quasi-particles'' consisting of the ``bare'' quarks and surrounding ``clouds'', which diagrammatically correspond to these insertions, are the ``constituent quarks'' which in the NQM (naive quark model) are the building blocks of all hadrons.

In strong coupling the clouds are represented by the ``Shrunk'' loops with the doubled $\bar{u}u$ or $\bar{d}d$  lines making up all the above side trees. In this limit and for all TPF's, the parts of the u and $\bar{d}$ quark lines which are not paired up in the above loops have to be glued together along the $u\bar{d}$ ``trunk'' connecting the points x and y. When $ l=|x-y|\to{\infty}$the wave function(al)s generated at intermediate times are those of the corresponding ground state mesons. All these mesons have the side chains which generated the masses of the individual quarks and which are responsible also for the corresponding clouds. Thus we expect all ground state hadrons to have in this limit, and also in general, similar extended structures. 

\begin{figure}[htbp]
	\centering
		\includegraphics[width=.25\textwidth]{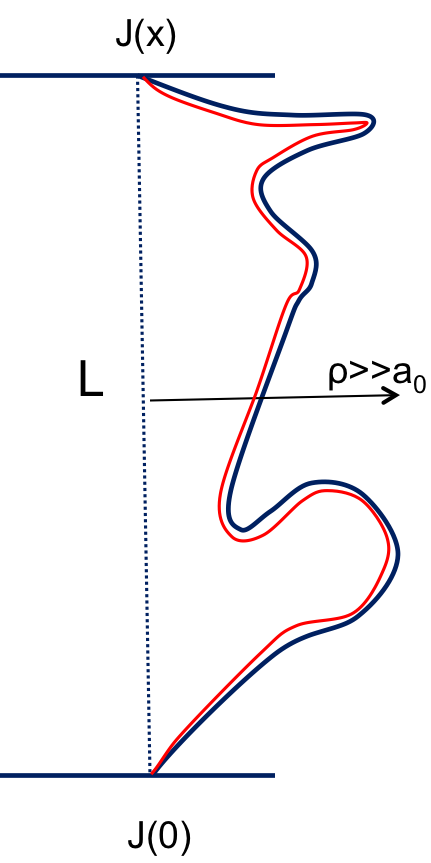}
		\caption{A self retracing pair of fermionic paths contributing to a $J_I(0) J_I(x)$ TPF. For the special case of the pseudo-scalar TPF's all these self retracing paths make a positive contribution. Consequently such paths can meander in the ``transverse'' spatial directions far outside the nominal strip of width $a_0$ indicating the extent of the confined wavefunction in Fig.~\ref{slide14}. Note that for simplicity we omited the side-branches and indicated the trunk namely the $\bar{u}d$ self retracing portion only}
		\label{fig:slide17}
\end{figure}

The key difference between the pseudo-scalar two point function and the rest is the positivity and coherence of the contributions of the various paths. Our basic conjecture is that this coherence allows extending (for the pseudo-scalar two point function) the ``trunk'' made of the original SR $\bar{d}u$ paths, along with accompanying side-chains/trees of doubled $\bar{u}u$ and $\bar{d}d$ lines, from the initial point x to the final distant point y and meander away from the original strip of width $a_0$ in Fig.~\ref{fig:slide17} without exponential suppression. The resulting ``Goldstone'' mass-less pions will then have a spatial structure similar to that of all the other mesons resolving the $\rho-\pi$ puzzle along the lines originally suggested in \cite{Pagles}.

 Each of the self retracing $\bar{u}u $ or $\bar{d}d$ side chain/tree consisting of $n_i$ doubled links, is a scalar insertion. Recalling the discussion of the scalar TPF above we find that any such insertion contributes a factor of $(-b^2)^{n_i}$. Let the total number of doubled links in the side trees be $\Sigma{n_i}=S$ and that in the trunk be $N-S$, so that N is the total number of links. Let the number of such paths which connect $x=(0,0,0,0) $ and $y=(L\delta,0,0,0)$ be $Q(N,S,L)$. Then $ F_{P}(l) =\Sigma_N b^{2N}\Sigma_S{Q(N,S,L)}.$ $(-1)^S $ is the contribution to the pseudoscalar TPF. We are presently attempting \cite{Berkowitz} to show that the inner sum yields a positive coefficient for every N.

  Assuming that our arguments or those of earlier authors indeed lead to massless pions or SXSB in the strong coupling limit the question of wether the result can be extended to the full theory is of paramount importance. For this to be the case no phase transitions restoring the axial symmetry should occur as the lattice spacing and corresponding coupling decrease and we approach the continuum limit. Since confinement readily manifests on the lattice in strong coupling \cite{Creutz} the analogue problem of showing that no deconfining phase transition occurs en-route to the continuum so that confinement holds also in the full theory is the stronger version of the millennium problem and phrasing the problem in this language may then not be so useful.

 However in the present case we can use point A above to directly argue that if pions are massless in the strong coupling limit, then they stay massless for  finite $g$ values.

 The increase of the energies/masses of bound states with increasing coupling is generic in vectorial theories. Thus consider two opposite abelian charges $q$ and $-q$ or a fermion and antifermion of opposite non-abelian charges belonging in representations $D(G)$ and $\bar{D}(G)$. Finite effects remaining after renormalizations smear the two point charges into densities $\rho_1(r);\rho_2(r)$ . The Coulombic energy is then the sum of the positive self energies $g^2[\rho_1. K.\rho_1]/2+ [\rho_2.K.\rho_2]/2$ and the mutual attractive interaction energy:$-[g^2\rho_1.K.\rho_2]$ The positive Coulomb kernel K\footnote{
The masses in interacting theories do ``run'' as different portions of the self mass due say to the Coulombic gauge interactions manifest for different distance scales though the effect is far larger for quarks than for electrons. The above positivity transcends these issues and holds separately for each cutoff in momentum space say. We can replace the momentum space Coulomb kernel  $K(q)=g^2/q^2$  by a cut-off version $K(q) .\Theta(q_0^2-q^2)$
\label{footnote:scale dependence}
}
ensures-via the Schwartz (or triangle) inequality- the positivity and monotonic increase of energy with the coupling $g$.

  Note that for {\it scalar} exchange the situation is dramatically different as now also the self interactions of $\rho_1(r)$ and separately of $\rho_2(r)$ are  attractive. In this case we do not have the Vafa Witten bound and the scalar attraction will bind the fermions below the sum of their bare masses- and conceivably even bellow zero so that the resulting mesonic system is Tachyonic even for massive scalars and corresponding short range interactions if the Yukawa coupling $g_Y$ is strong enough. This point is of crucial importance as having fundamental scalars at some scale will most clearly evade difficulties with obtaining the Higgs as a light bound state in a theory with a very high compositeness scale. This though is perfectly in line with our claim that it is difficult to obtain a composite $0^+$ 125 GeV particle in {\it purely} vectorial theories. If this were the route for making our light 125 GeV Higgs composite then we still would need fundamental scalar at some scale!

It is amusing to note that the ``Fatal Scalar Attraction'' most clearly manifests when we have ensembles of particles say quarks of mass M with \# density n. Let $N(eff)=n/{[(m_s)^3]}$ be the number of particles within the range $\lambda_s=(m_s)^{-1}$ of the scalar attraction. Since any of these particle is attracted by the rest with a pair-wise potential $g_Y^2\exp{-[m_s.r]}/r \sim{g_Y^2m_s}$ the system becomes unstable when $N(eff)g_Y^2m_s\ge{M}$.
The $\lambda\phi(scalar)^4$ repulsive self interaction energy density of the large scalar ``fields'' ( aka the potentials) which coherently builds up as the sum of the scalar potentials in the dense region in question tends to avoid a true vacuum collapse as $\phi\rightarrow{\infty}$.

 More formally consider the (logarithmic) g derivative: $g d/{dg}[\langle \Psi^0|H|\Psi^0\rangle]$ of the mass/energy of the light pseudoscalar particle written as the expectation value of the hamiltonian in $\Psi^0$- the corresponding eigenwave functional. The latter includes multi gluon and multi-$\bar{q} q$ Fock space states and can be expressed as a coherent sum of lattice configurations. The variational principle implies that:
 $\langle\delta (\Psi^0)|H|\Psi^0\rangle =\langle \Psi^0|H|\delta(\Psi^0)\rangle= 0$ so that we need to take the derivative inside the expectation value of the hamiltonian $\int{d^3x [g^2E^2+B^2/{g^2} + g J.A]}$. The $J$ above is the vector part of the color current and the $J_0.A_0$ Coulombic part emerges from the $g^2E^2$ and the Gauss constraints: $Div E= gJ_0= g[\psi^+\psi]$. The derivative is $\int{d^3x [2(g^2E^2-B^2/{g^2}) + g J.A]}$. The expectation value of the last expression is positive since in the bound state the electric term $g^2 E^2$ dominates over the magnetic $B^2/{g^2}$ term. Also the constraint implies that $E$ increases with the coupling $g$.

The prominence of the electric part of the energy is implicit in the previous heuristic argument utilizing the electric Coulomb rather than the magnetic vector current interactions. It also clearly manifests in the chromoelectric flux tube which is assumed to extend between the bound Fermion and anti-fermion in the fundamental representation. This is very different from the vacuum where the magnetic fluctuations dominate.

 The basic feature that masses of the fermionic bound states always increase when the gauge coupling is turned on holds for fermions with any bare masses $m^0(a), m^0(b)$ in any spin parity channel and in particular to the lightest pseudo-scalar states. The above heuristic arguments neatly tie in with the upper bound of Vafa and Witten \cite{Vaffa} on Fermionic propagators in the gauge field backgrounds which yields the following inequalities between the masses of the lightest pseudoscalars $M^0_{P}(a,b)$ and the {\it bare} masses of the quarks they are made of
\footnote{
 These inequalities can be directly proven only for the PS case with different flavors ( to avoid the flavor disconnected contributions) yet equal masses (so as to have the positive definite $S^+S$ integrand). Neglecting the $\bar{Q}Q$ annihilation diagrams which is justified for the heavy b and c quarks we have the upper bounds on their bare masses: $M(\eta_c)\ge{2m^0(c)}, M(\eta_b)\ge{2m^0(b)}$ which are consistent with estimates by A. Manohar in\cite{PDG} where $m^0(c)$ and $m^0(b)$ lie in the intervals of $1$-$1.4$GeV and $4$-$4.4$GeV respectively.  The interflavor mass inequalities for ground state pseudoscalars proven in \cite{Witten} and motivated by variational arguments \cite{Nussinov1} and \cite{N&L} allow extending to unequal bare masses:
 $M^0_{P}(a,b)\ge 1/2[M^0_{P}(a,a) +  M^0_{P}(b,b)]\ge {[m^0(a)+ m^0(b)]}$
} 
\begin{equation}
              M^0_{P}(a,b)\ge{[m^0(a)+ m^0(b]}
\label{equation:no binding bellow bare masses}
\end{equation}

 The number of paths or of closed loops with N links encountered in treating statistical mechanics models on lattices often increases asymptotically  as $c^N . N^{-\alpha}$. If the contribution of any single path or loop with N links is $\sim{K^N}$, where K is a constant, then these high temperature expansions have a finite radius of convergence of $K=c^{-1}$ with $K=J/T$ the ratio of a, typically nearest neighbor, coupling and the temperature. The singularity in the complex K plane at $K=c^{-1}$  then fixes the ``critical temperature'' or `` critical coupling'' and $\alpha$, the ``critical exponent'', fixes the type of the singularity, generically a branch cut of the form $(1-cK)^{\alpha}$.

In the field theory analogs of these models we have cK equal (or extremely close) to 1 and the theory should be such that this is a second order phase transition allowing long correlation length and hence generating a smooth continuum limit. Since N>L, a common $(cK)^L$ factor multiplying the $F_I(|x-y|)$ of interest should be one to avoid the exponential falloff  .

The anti-commutation of fermionic fields and the ensuing Pauli exclusion principle tend to regulate or soften the singular behavior in the case of  fermions. Weak coupling expansions in interacting field theories are asymptotic. The number of nth order diagrams  of diagrams increases like $n!$ in $n$th order which makes the bosonic perturbation series diverge in order $1/{g^2}$. While the counting of diagrams is the same in the fermionic case considerations of how the Pauli principle modifies the original Dyson argument for the singularity delaying the onset of divergences.

Also in the present case the fermionic nature does not directly decrease the number of fermionic paths that we sum over but manifests much more subtly via the pattern of signs due to the algebra of the $\gamma_{\mu}$ matrices. Thus consider some general fermionic path P connecting $x$ and $y$ which enters a point r along the x direction say via the link $(n,\mu_x)$ and leaves it along the $y$ direction via $(n+\mu_x,\mu_y)$. We can distort P to P' by one plaquette so as to arrive the opposite corner of the plaquette via $(n,\mu_y)$ and leaves it via $(n+\mu_y,\mu_x)$. In both cases the original path $P$ and the distorted path $P$ start from the same point n and arrive in two steps at the same $n+\mu_x+\mu_y$. Because $\gamma_x$ and $\gamma_y$ anti-commute we can do this change at the expense of a - sign. We can repeat these steps until we arrive at some standard path of the same length and which still connects x and y. Once we properly understand how the correct propagator $S^0_{m=0} (x,y)$ of a free massless lattice Dirac fermions \cite {Carpenter} arises we will be able to properly calibrate the all important parameter b.

 The key element of our conjecture is the absence of  the non-abelian (Wilson loop type phases) for the SRP's and for the augmented set with shrunk side loops described above {\it and} the constant coherent sign of the remaining spinorial factor for all paths of different \# of links in the case of the strictly SR, trunk like paths (and presumably also for the above augmented set). This essentially neutralizes the mechanism for dynamically generating exponentially falling $F_P(|x-y|)$ described in Sec. 5 above. In particular the SRP trunk- like path can meander  far away from the strip propagate this very far distance $l=|x-y|$

The only remaining potential source of exponential fall in this case is the actual value of the hopping amplitude $\sim{b^2}$ playing the role of K above. In order to be able to show that this is not the case we {\it must} use the fact that the fermions/ quarks involved have vanishing bare masses. Indeed if fermions $F_a$ and $F_b$ had non-zero bare masses $m^0(a)$ and $m^0(b)$ then the Vafa Witten bound mentioned above implies that the  lowest mass of the full interacting system would be $[m^0(a)+m^0(b)]$  and only scalar attractive interactions can reduce it any further.

 We will return to this issues and also to a more careful analysis of cases when the paths are not exactly self-retracing and their increased number is compensated by the increased flux threading the loop and the ensuing phases.

 Instead of using a lattice we can discretize time only. The path P is then made of N links each of length $l_i$  along the unit vector $n_i$ and likewise for the return path P'. The spinor factor S(P,P') associated with the closed PP' loop carries then the product :$\Pi (n_i.n'_i)$ which becomes unity for sections section in P and P' that are parallel. This more flexible form appears to be useful when the paths are slightly separated from each other, which one must consider in the continuum limit.
 \begin{figure}[htbp]
	\centering
 		\includegraphics[height=.33\textheight]{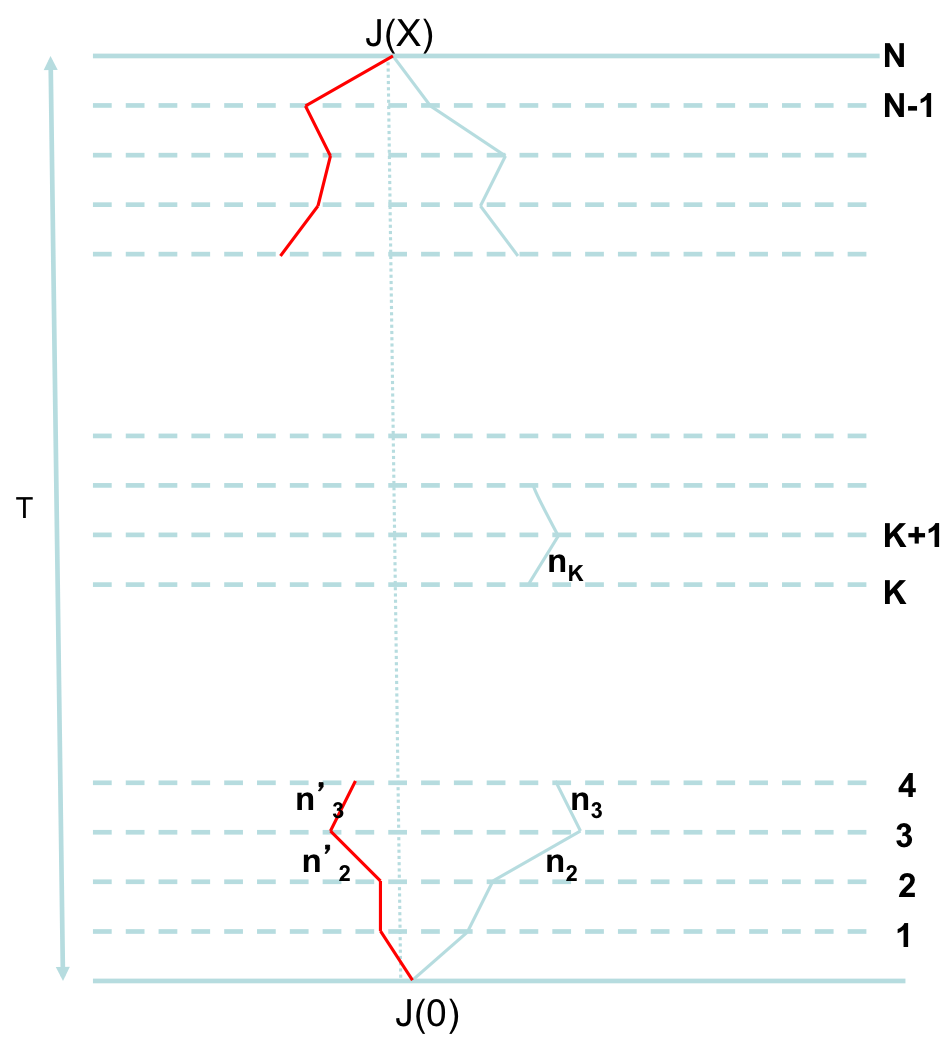}
		\caption{the loos PP' divided into time slices where the fermion anti-fermion travel along short sections along the directions n and n' respectively}
		\label{slide16}
\end{figure}

\bigskip

\section{Experimental LHC Information and our Assumptions Pertinent to a Composite H(125) State}

In the the previous two sections we emphasized the qualitative difference between the scalar and the pseudo-scalar TPF's  and consequently between the massless lowest pseudo-scalars and the massive scalars in vectorial gauge theories with massless Dirac fermions. In the following two sections we explain how this impacts the LHC discovered ``Higgs'' particle and why it makes many composite Higgs models unlikely. For clarity and completeness we reiterate next in some detail the assumptions we need, explain why these are reasonable assumption and finally speculate if (and how) breaking some of the assumptions will indeed allow reasonable composite Higgs models.

 LHC searched mainly SUSY particles and string/brane inspired extra $U(1)$ gauge bosons ($ Z'$ s) or KK (Kaluza Klein) recurrences associated with extra dimensions. While searches of ``Technicolor like'' new confining gauge theories were less dedicated the bounds on new Z' and/or KK particles do in fact provide bounds on the scale of a putative new confining gauge theory or the ``compositeness'' scale of any theory which dynamically breaks the E.W. symmetry. Thus consider heavy $X^0$, or $X^+$  states in a new gauge theory which generates a composite H(0). It contains fermions such that the spontaneously generated V.e.V. of their bilinears breaks the E.W symmetry. Using the `techni-$\rho$' decay into two technipions as a qualitative guide, we expect the $X^0$ or $X^+$ to decay into WW, or WZ  because the N.G. Techni-pions are the longitudinal modes of the W and Z bosons. These S.M. gauge bosons then decay with reasonable branchings to leptons and searches for such final states with large invariant masses strongly limit such $X^0 , X^+ $ particles . The X decay widths to the above, presumably dominant two E.W. boson decay channels, are suppressed by $\alpha_{weak}^2$ factors and are much smaller than $m(X^0) ,m(X^+)$ so that we can search for bumps in the invariant mass of the multi-lepton combinations in the final states. The strong O(TeV) bounds on the KK recurrences and/or extra $Z's$ or right handed $W'$ s then suggest similar strong bounds on $m(X^0)$ and $m(X^+)$ and on the scale $\mu$ of the theory. Still we should emphasize that if none of the new fermions carries color and the purely E.W production of this new sector particles may be small enough to weaken the bounds to roughly $\mu\ge{1/2TeV-1 TeV}$

 So far we focused on confining theories where the SXSB occurs very much in the same way as in QCD say techni-color and it's extended versions.   

The SXSB responsible for breaking the E.W. symmetry could occur in a non-confining theory. In this case  we have a new family of unconfined massive fermions the lightest of which should be stable. These masses and the scale of the SXSB generating them should also be high enough so as to evade being discovered in LHC searches. The fermions involved which carry SU(2)$_L\times$U(1) quantum numbers could be massive to start with.  The EW symmetry then breaks at a much lower scale via a slight ``vacuum misalignment''. The basic Vafa-Witten bound implies that without introducing new dynamics beyond a purely vectorial underlying theory all particles have masses exceeding $\sim{2m(Q)}\ge$TeV.

This then motivates
\begin{description} 
	\item{{\bf Assumption I:}}  the new scale of the dynamics which generates the composite Higgs is higher than O (1/2-1) TeV.
\end{description}
The distinction between the scalar $J^P=0^+$ and pseudo-scalar $J^P=0^-$ states is crucial  for our discussion. LHC experiments can tell us the parity $(P)$ and not only the $CP$ of the  $125 GeV$ state. Thus:
\begin{description}
\item[i)] The two photon decay discovery channel implies that H has even charge conjugation C so that even CP implies even P. 
\item[ii)] The analysis of the experimental angular correlations between the decay planes of the two leptons from $H\rightarrow{ZZ^*}$ decays suggests a $O^+$ state. 
\end{description}

 In principle parity can be unambiguously measured in $H$ production which proceeds mainly via two gluons and the top triangle. {\it if} both gluons have sufficient transverse momenta to define the planes of the incident protons and recoiling quark jets , then excess of parallel (perpendicular) planes indicates $0^{+(-)}H(125)$. Such events are rare relative to those with low transverse momenta, and much fewer than those originating from Vector boson fusion where a similar correlation might be expected.

The basic question which needs to be addressed is whether we can infer parity in the first place in the presence of the parity violating weak interactions.

 To this end we recall that the first $H\rightarrow{\gamma+\gamma}$ decay is loop induced and involves circulating charged particles which in the SM are  the top quark generated by the QCD process of $g+g\rightarrow\bar{t}t$ and the charged $W$ loop. In beyond SM extensions there could be some small corrections due to loops of heavier particles. The top strongly couples to the Higgs and the first QCD process dominates the rate. Apart from weak EW two loop radiative corrections no parity violating physics involved here. As for the second $W$ loop contribution - there are no fermion loops involved here and the Xiral nature of the $SU(2)_L$ weak coupling is irrelevant .

  We can indeed infer from (ii) the positive parity of the Higgs. The very observability of H to VV decays decays indicates that these are not loop induced but are-as predicted in the SM- directly generated by the $g_{Weak}^2 H^+HVV$ coupling when we replace one of the Higgs fields by its VeV:
$\langle H\rangle = v$ and this coupling clearly indicates a scalar H. Indeed a pseudo-scalar H cannot have a non-vanishing VeV since vectorial gauge theories do not spontaneously break parity\cite{Vafa3}. 

 In view of the above we will make our 
\begin{description}
 \item[\bf Assumption II:] The parity of $H(125)$  is well defined and positive.  
\end{description}

It should be emphasized that a confining dynamics at a higher scale with a strong {\it Xiral} gauge theory component could allow the pseudo-scalar NG boson to have a VeV and thus masquerade as an ordinary scalar Higgs in the LHC experiment at the scales probed so far\cite{Sundrum}. To exclude this possibility and to make sure the putative theory which is presumed to give rise to a composite H(125) is parity conserving we make our third and most important
\begin{description}
 \item[\bf Assumption III:] 	The underlying gauge dynamics which produces the $H(125) $ $0^+$ particle as a composite, is purely vectorial.
\end{description}
  The great advantage (or draw-back- depending on one's point of view\ldots) of such theories is that the path integral expressing all the n point euclidean functions and in particular the TPFs -two point function which determine the spectrum of the theory has a {\it positive} measure. This along with lattice discretization allows Monte-Carlo simulations which after $\sim{40}$ years reproduce most of the observed spectrum of QCD and many other features.

 In the present context this positivity is important because it allows us to derive inequalities between various two point functions which translate to mass inequalities. The latter are then used, in various reincarnations, to show that the lightest scalars are significantly heavier than the lightest pseudo-scalars with the difference being of order of $\mu$, the scale of the theory.

 Most  SUSY and Xiral models which have been the focus of interest in recent years do {\it not} have a positive path integral measure and did not , so far, allowed a consistent lattice discretization. This diminishes our ability to make dynamical predictions in these theories in a manner similar to what we attempted above, though extensive research in SUSY theories led to a rich set of results and conjectures by Seiberg, Witten and Maldacena. Also the Xiral $SU(2)_L$ gauge group is a key part of the perturbative S.M. Notwithstanding our technical (?) difficulties nature may use also strongly coupled variants of such theories.

 Since all the above depends on the positivity of the path integral measure we cannot argue that the scalar mass is high when the positivity fails, e.g in Xiral gauge models. Note however that the Xirality condition $\gamma_5\psi=+/-\psi $ preventing the pairing of the eigenvalues of the Dirac operator into complex conjugate pairs, as required in order to ensure the positivity of the determinantal factors in the measure, makes $\bar\psi H\psi$ and
 $\bar\psi H\gamma_5\psi$  identical. There is then no distinction between scalars and pseudo-scalars and the lowest physical state in the $\bar{F}_aF_b$ channel is likely to be an equal mixture of both. Alternatively we should have a Xiral multiplet with almost degenerate $0^+$ and $0^-$ states for which there is no evidence and LHC data favor a well defined, even parity, isolated state. Thus while Xiral dynamics may produce light scalars these may not be the $0^+$ state at $125 GeV$.

  Implicit in most discussions of  the $125 GeV$ particle is the assumption that it is connected with the EW symmetry breaking. This implies that if composite it is made of a fermion $F_a$ and anti-fermion $F_b$ which carry $SU(2)_LXU(1)$ quantum number. Because of the spontaneous breaking of EW symmetry the composite $0^+$ state could mix with a pure ``Glue-ball like'' $0^+$ state, a composite of the gauge bosons of the underlying confining group and such a mixing could in principle reduce it's energy. While this is a remote possibility\footnote{
 As we noted above the non-vanishing VeV of $F^2_{\mu,\nu}$ in confining gauge theories suggests that the lowest glueball mass is high. While we should not take the case of QCD too literally, both lattice calculations and the lack of experimental evidence suggest that $M^0(gb)\ge{1.6 GeV}\sim 4\mu$ with $\mu\sim{400 MeV}$ so that a scalar of mass $\sim{\mu}$ is unlikely to strongly mix with it. In the case of the ninth pseudoscalar that strong mixing dramatically increased the mass of $\eta'$  to ~ $950 MeV$ a mass which stems from the axial U(1) anomaly breaking the conservation of the ninth axial current. In view of the heavy GB masses in the $0^+$ and in the $0^-$ channels we can (\cite{Nussinov2}), describe the propagation between the distant x and y in the scalar and pseudo-scalar TPF's as proceeding most of the time via fermionic loops connected by effective four fermion couplings generated via the exchange of the heavy GB states over relatively short distances. Summing the resulting geometric series in momentum space tends to yield higher masses than for the case when no annihilation is possible. Still the difference between the pseudoscalar case where the mixing definitely pushes up the mass and the scalar channel leaves open the possibility that the mixing with the pure gauge sector will lower the mass of the scalar.  
\label{footnote:mixing is of no use}
} we prevent any such mixing and the generation of any extra fermionic loops by adopting: 
\begin{description}
 \item[\bf Assumption IV:] 	The large $N_c$ limit can be used.  
\end{description}

 The large $N_{c}$ limit is very important in several ways: it allows confinement even of light dynamical fermions, it is implicit in the QCD mass inequalities and the proof by Coleman and Witten, \cite{Coleman3} of SXSB by using axial anomaly matching arguments and all other arguments for SXSB  including the one we presented in Sec VI above do rely on this limit.  

  To show (or argue) that the scalar is heavier than the pseudoscalar by the ``scale of the theory'' we need that the vectorial theory with the mass-less Fermions in question will indeed have a scale in the first place. Perturbative U.V.  QCD with massless (bare) quarks is, just like QED with massless electrons,  scale invariant. That non-perturbative effects generate in the pure Y.M. theory a scale a.k.a. ``mass gap'' is then our fifth explicit assumption:
\begin{description}
 \item[\bf Assumption V:] 	The strongly coupled Y.M. theory develops via non-perturbative effects a mass gap. 
\end{description}

 The mass gap serves as an infrared cutoff and it's inverse provides the finite correlation length for the vacuum gauge fields.

 While confinement is conceivably derivable from Assumption V we make it a further explicit  
\begin{description}
 \item[\bf Assumption V':]The Y.M. theory confines the fermions $F_a,\bar{F}_b$.
\end{description}
Specifically the large planar Wilson loops satisfy the area law. Furthermore we assume that the confinement arises via the ``inverse Meissner effect'' that is by having strong chromo-magnetic B field fluctuations in the vacuum. 

 The substantial difference between the mass $\sim{\mu}$ of the scalar and the massless pseudo-scalars can then be shown by:
\begin{description}
	\item[a)] ``proving'' that assumptions V and V' above cause all particles in the theory to have generic masses of order $\mu$ 
	\item[b)] recalling that thanks to the SXSB the pseudoscalars in the  $F_a,\bar{F}_b$ channel remain massless Goldstone bosons. 
\end{description}

  We attempted to motivate point (a) in Sec V above and in sec VI speculated on the mechanism which allow the PS TPF to evade the generic 
$F(|x-y|)\sim\exp{-[\mu|x-y|]}$ asymptotic fall off of all other TPF's.

 We are sure that notwithstanding the simplistic and crude nature of the arguments of Sec V the result claimed there is indeed correct and can be rigorously proven. This is certainly not the case for our novel conjecture  that a specific slightly augmented set of self-retracing paths is responsible for the fact that the PS TPF $F_P(|x-y|)$ has no exponential fall. Our arguments there are at best incomplete but we hope to improve those soon \cite{Berkowitz}.

If correct this conjecture would provide new useful insight into the dynamics of SXSB  and has applications in QCD phenomenology. We should however re-emphasize that it is {\it not} essential for our main point that vectorial gauge theories at a high scale are unlikely to produce a light, scalar, composite Higgs.  

\bigskip

\section{ Limits on Dynamic Composite Models and how these can be Alleviated by relaxing some of our assumptions .}
 
  In discussing possible implications of the present paper for composite Higgs models based on vectorial underlying gauge interactions we will focus on two main classes of models which broadly speaking we refer to as (extended) Technicolor and generalized ``little Higgs'' type models. 

  The idea that spontaneous breaking of conformal invariance yields a low mass pseudo Goldstone dilaton has been recently revived (for details and many references see \cite{Yamawaki}). Indeed apart from the effects of the Higgs the S.M. with massless gauge bosons and massless fermions is conformal \cite{Bardeen} and in certain models with scalars $\phi_a$ with dimension-less couplings respecting conformal invariance \cite{Bander} the latter can spontaneously break yielding a massless (or a very light) dilaton. Dilaton mass and couplings were discussed in e.g. \cite{Goldberger} and an analysis in the context of light Higgs \cite{Appelquist,Bellazzini,Chacko1} was recently done. All these are of great interest but the question addressed are different from ours, namely whether $\bar{F}-F$ composite light scalars can naturally account for the H(125) scalar.

 It may seem that there is no difficulty in achieving this. In particular in extended, ``Walking'', Technicolor the number of technicolors and of flavors of the technifermions can be tuned so as to almost yields an infrared fixed point (with vanishing beta function) at $g=g^*$ and a corresponding conformal theory. However it is argued that if at $g=g^*-\epsilon$ a techniquark antitechniquark $\langle \bar Q Q \rangle \sim v^3$ condensate with $v \sim 250 GeV$ forms, providing the requisite E.W. SSB, it spontaneously breaks also the almost exact conformal invariance and a scalar pseudo N.G dilaton may arise.

A proper choice of fermionic representations can radically change the theory and in particular can make it almost conformal. But as noted in the preceding section if the scale of the theory defined by its lightest non-Goldstone particles is high, then regardless of any fancy theoretical setting, the difficulty of generating Light scalars may persist.

In addressing the above models and the other class bellow the following general considerations are useful. 
 In gauge theories with fermions but with no fundamental scalars we are much more limited in choosing spontaneous symmetry breaking (SSB) patterns than in theories with fundamental scalars-including the minimal S.M. Higgs, where tuning the tachyonic mass and the quartic coupling fix the Higgs mass and VeV.

 No matter how hard we try to ``induce'' the pure vectorial gauge theories to produce a desired pattern of SSB the theory may fail to oblige. The SSB of non-abelian grand unified gauge theory or of other gauge theories to the gauge group(s) of the standard model, directly or via some intermediate stages and or breaking of L-R symmetric models to he diagonal or left parts are relevant examples. The theorem \cite{Elitzur} that local gauge invariance does not spontaneously break, was overcome by the complementarity approach \cite{Fradkin}.
 However, the control parameters available in purely vectorial theories namely the choice of the fermionic representations, may be insufficient to achieve the desired S.B pattern. This difficulty readily overcome by adding  generalized Higgs fields $\phi_i$. With appropriate $m^2_{i,j}\phi_i^*\phi_j$ quadratic terms and a quartic $\lambda_{i,j,k,l}\phi_{i}^*\phi_j^{*}\phi_k\phi_l$ interactions subject to the requirement of symmetry under the Gauged and global ``Flavor'' symmetries, we can achieve almost any desired pattern of symmetry breaking.\footnote{
Some general positivity constraint are needed to avoid ``run-away'' directions in $\phi$ space where $V(\phi) \rightarrow-\infty$. If however the desired  manifold of local minimum of $V(\phi)$ is separated from these ``run-away'' regions by a sufficiently extended and high barrier in $\phi$ space the metastable vacuum can tunnel out only on super cosmological times.
\label{footnote:Higgs potential positivity constraints} 
}

  The above considerations are most readily applied to Dynamical E.W. breaking QCD like schemes such as Technicolor in which the SXSB directly yields  the massless pseudoscalar Goldstone bosons eaten by the W and Z and where light Higgs were not expected until the more recent Techni-dilaton constructs. The applicability to the rather different compositeness scenarios such as ``Little Higgs'' and related models \cite{Schmaltz,Arkani-Hamed} is not as obvious but we still believe that also here it may be difficult to menage without introducing some fundamental scalars and/or some axial gauge interactions.

Models of this type were originally introduced by Kaplan and Georgi and collaborators \cite{Kaplan2,Kaplan3,Georgi1,Georgi2}. A key element is the separation between a high ``Compositeness'' scale where an original strong gauged group and an extended flavor symmetry of the confined fermions partially break, and the E.W symmetry breaking at the ``low'' $(v_0\sim 250 GeV)$ which is dynamically generated via some slight ``vaccuum misalignment'' induced by some other competing gauge interaction. The lightness of the EW and Higgs masses are protected by having {\it all} light bosons PNGB's initially pseudo-scalar essentially massless bound state in the underlying theory at a higher scale.

  However that as long as all the gauge interactions those at the higher scale and the ones which persist at the lower scale and the gauged flavor groups are are vectorial the whole arsenal of the QCD inequalities and the Vafa Witten Bound  are still applicable. More specifically we note that:

\vspace{1ex}
{\bf i)} For any additional gauge group of a direct product form we need to multiply the measure by another $\exp(-[\int {d^4x[E'^2+B'^2]})$  positive factor and replace $D(A_{\mu})$ by $D(A_{\mu},A'_{\mu})$ the Dirac operator in the presence of both gauge fields (see e.g the proof in \cite{Nussinov1} that $m_{\pi^+}-m_{\pi^-}\ge{0}$). Thus having any direct product structure of various gauge groups at different scales and strengths does not circumvent the difficulty of having light composite scalars made of Fermion and anti-fermions in vectorial gauge theories.\footnote{
 A more general situation where the gauge groups are not exactly of a direct product form but partially ``conflict'' by having some generators which do not exactly commute is unfortunately not possible in the conventional $UV$ completion using renormalizable non-abelian gauge interactions. We must close the algebra by completing the system to one or several mutually commuting but otherwise complete gauge groups. 
\label{footnote:not X product form GGs}
}

\vspace{1ex}
{\bf ii)} If in the process of breaking the symmetry at the high scale one generates massive fermions -which are practically elementary at the lower scale - the Vafa-Witten bound forbids binding them into light composites by vectorial interactions.

\vspace{1ex}
{\bf iii)} When a large symmetry group $G$ say $SU(N+M)$ breaks down to $G_1XG_2$ say say $SU(N)XSU(M)XU(1)$ at some high scale $\Lambda^*$ the $2NM$ off-diagonal, broken generators correspond to massive vector bosons. Their exchange mixes fermions which transform under one of the unbroken groups or the other and vice versa can be represented via effective four fermi couplings-and  in principle might generate some low lying scalar states. However failing by the arguments given above to achieve such binding in the original unbroken big group with $(M+N)^2-1$ massless bosons one may wonder why having less than half of them massive with corresponding short range interactions will enable doing it.

 Repeating the heuristic argument that, due to the repulsive self interactions, turning on a gauge interaction and exchanging vector bosons between a fermion $F_a$ and an anti-fermion $\bar{F}_b$ of bare masses $m^0(a)$ and $m^0(b$ {\it raise} the mass of the system above $m^0(a)+m^0(b)$ 
for the case of massive vector bosons is very instructive. We find that the net effect of self and mutual interactions is still to raise the energy of the system above the bare mass threshold- namely $\Delta=m^0(a,b)-[m^0(a)+m^0(b)]$ is still positive but due to weakened attraction less so than in the case of massless gauge bosons.

 If gauge vector bosons mass were generated by a Higgs mechanism there is a very suggestive interpretation of this result: along-side the vectorial exchange we exchange also the massless scalars which as noted above are attractive and in general they may well break the Vafa Witten bound and allow arbitrarily strong binding of the fermion anti-fermion system. however if the effect is subsumed by having just massive vectors this bound and all the other difficulties we pointed out may persist.

To motivate this in a different way let us introduce the masses of the vector mesons by hand by adding to the euclidean path integral YM action the term $\int{d^4x M_V^2B^2_{\mu}(x)}$. This is problematic from the point of view of gauge invariance and renormalizability. Still it is less problematic than the local four Fermi interactions and in the spirit of effective Lagrangian may be fine. For our purpose we note that adding this term does not affect the measure positivity and all our results may apply here too.\footnote{
 Since we can put the vectorial gauge theory - also in the presence of elementary scalars- on the lattice we have a nice way of transiting between different effective descriptions at different scales. This is done by coarsening the lattice and modifying a-la Symanzick's the original Wilson action to improved versions.
}

\vspace{1ex}
{\bf iv)} 
When $N_F \sim {N_c}\rightarrow{\infty} $ {\it all} planar diagrams, also those with many fermionic loop,\cite{Veneziano} need to  be retained.
 Even if not at or near the conformal point with the infrared zero of $\beta$ function- the suggested  dilaton ``escape route'' it may allow certain composite scenarios to avoid the basic difficulty pointed here. Thus having many $\sim N_F^2\;$ NG bosons one may wonder if the effective low anergy Xiral Lagrangian describing them can in fact generate a light scalar- the $125 GeV$ Higgs particle in the case at hand - as a bound state of some combination of NGB's pairs. While this may happen we believe that it is rather unlikely in general and for the case at hand in particular. It is hard to conceive of the 125 Higgs particle being a bound state coupling to a pair of $W$s without some anomaly near the WW threshold and without an enhanced $H\rightarrow{WW}$ decay.  At large NF we are likely to have many more massless NG particles for which there is no indication at the LHC\cite{Shrock}.

 In general large departures from pure vectorial models---such as a purely Xiral gauge theory or a large $\theta \tilde{F}F$ term do break as shown above  measure positivity. However unlike analyticity, positivity and the various inequalities derived using it are robust- and may not be affected by adding small ``positivity violating'' terms to the Lagrangian. Thus consider the modifications (a) $[E^2+B^2]/2+i\theta(E.B)$ in the pure gauge part of the Lagrangian or (b) $\gamma_{\mu}+\epsilon.\gamma_\mu\gamma_{5}$ in the fermionic part. If $\theta$ is small then in order to generate  significant oscillations or phase in $\int{d^4x [(E^2+B^2)/2+i\theta(E.B)]}$ we need that the integral of $\theta(E.B)$ over some region R of interest be significant say $\sim{\pi}$ but than the integral of $[E^2+B^2]$ over this region exceeds $\pi/{\theta}$ dramatically suppressing it's contribution by more than $\exp{[-\pi/{\theta}]}$.

 The gauge field configurations maximizing the oscillatory effects are the pure (anti-)instanton configurations for which the field is (anti) self-dual i.e.
$ E=\pm B$ in which case we have precisely the above suppression which is indeed minimal when $\theta$ has it's maximal value $\theta=\pi$.  Likewise the effect of adding an axial current to the interacting vector current becomes maximal for chiral theories when $\epsilon=1$ . Thus there seems to be an annoying(!) complementarity that when the ``elegant'' purely Xiral, purely instanton physics prevails positivity is maximally violated.   

\section{Summary and further comments}.

 To better understand dynamical theories which potentially generate the observed 125 GeV $0^+$ as a bound state, we revisited confining vectorial theories with massless quarks/fermions. Our main objective was to understand by studying the Euclidean two point functions of bilinears of the above fermions at large separations why the lightest pseudo-scalars are massless or very light whereas the $0^+$ scalars have the O(TeV) mass scale of the theory. Under a very specific set of assumptions including that the scale of the new compositeness physics is high, O(TeV), we therefore find that vectorial gauge theories are unable or unlikely to produce scalars as light as the $125$ GeV $0^+$ as a fermion-antifermion bound state.

Within string  theory and string/brane inspired extra-dimensions the question posed here wether the H(125) particle elementary or composite may be mute. We can have elementary, weakly coupled, ``electrical'' charges and highly composite magnetic monopoles (consisting of $ \sim 1/\alpha$ bosons \cite{Drukier}) with a reversed situation in a dual phase. Also in theories with extra dimensions compositeness may become relative and we could have various degrees of compositeness depending on the distance in the extra dimensions to some brane. Still we believe that the study we undertook here of the prospects of composite H(125) particle in local vectorial gauge field theories is worth-while. While it suggests that many such models will not work trying to evade our ``pseudo no-go theorem'' may restrict the model building in a useful way and help point out to the correct alternative- be it an elementary or composite H(125) particle.

 We have actually hinted at some such possibility by noting that  a light fermion- light fermion (rather than fermion -anti-fermion) with the fermions transforming as real representations of the confining gauge group form a light {\it scalars}. related example involve $SU(3)$ \cite{Contino} or other \cite{Keren-Zur} flavor groups. Such schemes may require mixings between composite higher scale baryons and fundamental quarks\cite{Kaplan4}.

Finally, finding at the LHC a relatively light compositeness scale or near-by opposite-parity Higgs particle will clearly ameliorate all the issues brought up here.
 
The second part of our paper is devoted to a conjecture on a path integral mechanism for massless pions and SXSB in QCD and vectorial theories in general. 

We suggested that self-retracing paths and a slightly extended class of similar paths dominate the PS TPF and generate massless pseudoscalars in the $m^0(u)=m^0(d) = 0$ limit. Eventually this may lead to an independent proof of massless pions and SXSB in QCD and in vectorial theories in general. In the large $N_c$ limit adopted here the argument of \cite{Coleman3} using axial anomaly matching or equivalently the baryon meson mass inequality in large $N_c$, do prove that the pions are massless. Thus no proof of SXSB in this limit is needed and none was provided here. Our attempts to understand this using the long distance behavior of the pseudo-scalar two point function are non-the-less a useful complement. 

 There may be more direct connections between the various approaches to SXSB and massless pions. The axial anomaly is related to the peculiar kinematics where a massless pole is generated in the $\gamma_5$ channel when the light-like momenta of the on shell massless fermion and that of the antifermion are parallel. Both the Banks-Casher criterion for SXSB of having the zero modes of the Dirac operator behave as if they were those of a free 1 dim (where four momenta are always parallel!) and our conjectured  importance of the contribution to the asymptotic behavior of  pseudo-scalar two point function made by self retracing paths with parallel moving fermion and antifermion seem to be intimately related to this.    

While inspired by our desire to understand the scalar -pseudoscalar mass difference. is of interest in its own right and hopefully will be proved (or disproved)  by stronger mathematical physicists. It may also ''justify'' to some extent the phenomenology driven strong hyper-fine interactions introduced in the NQM. In this context we note that for u and d quarks (rather than u and $\bar{d}$) in the $0^+$ ( rather than the $0^-$!) channel the contribution of overlapping u and d paths in the path integral for the nucleon have a  consistently positive spinor factor. This ud spin singlet then forms the $\bar{3}$ color state with a  suppressed Wilson loop color factor, namely the relatively tightly bound diquark which is a key element of the NQM phenomenology.

\section*{Acknowledgements}

This work was being done over the last year and a half during which I benefited from discussions with numerous colleagues .I am indebted to K. Agashe, T. Banks, Z. Chacko, F. Englert, and T. Volansky for discussions. Comments made by C. Bernard H. Neuberger, I. Shamir, L.Susskind, B.Svetitski and N. Yamamoto were very helpful. I should note in particular the helpful discussions with W. Bardeen on spontaneously broken scale invariance, with A. Casher on all aspects of SXSB in QCD, with N.A. Hamed, Y. Nomura, L. Vecchi and R. Sundrum in particular on the Non-techni-color composite models, with  Z. Nussinov on condensed matter analogies, with M. Ogilvie on possible rigorous arguments, with A. Vainshtein on the ``$\rho-\pi$'' paradox,  with P. Bedaque, E. Berkowitz, T. Cohen A. Nicholson and S. Wallace on lattice fermionic propagators. I am greatly indebted to E. Berkowitz, T. Nussinov and Z. Nussinov for help with the paper, and to R. Shrock for closely connected joint work \cite{NSh} and for instigating the precursor to this work \cite{Shrock}. Finally a discussion with E. Witten made me more strongly realize the lack of rigor of this work. While further efforts inspired by this did not (as yet!)  add much to rigor they clarified the precise assumptions and conjecture made.

\end{document}